\documentclass[aps,prd,letterpaper,showkeys,showpacs,twocolumn,nofootinbib,preprintnumbers,longbibliography,unsortedaddress]{revtex4-1}

\usepackage[top=2.8cm, bottom=2.8cm, left=2.4cm, right=2.4cm]{geometry}
\usepackage[T1]{fontenc}     
\usepackage{amsmath,amssymb,lmodern,amsfonts} 
\usepackage{graphicx}
\usepackage{hyperref}
\usepackage{color}
\usepackage{natbib}
\usepackage[caption=false]{subfig}
\usepackage{geometry}
\usepackage{booktabs}
\usepackage{amstext}

\makeatletter

\providecommand{\tabularnewline}{\\}

\makeatother

\newcommand{\gsim}{\gtrsim}

\begin{document}

\title{Anisotropic $k$-essence}

\author{J. Bayron Orjuela-Quintana}
\email{john.orjuela@correounivalle.edu.co}
\author{C\'esar A. Valenzuela-Toledo}
\email{cesar.valenzuela@correounivalle.edu.co}
\affiliation{Departamento  de  F\'isica,  Universidad  del Valle, \\ Ciudad  Universitaria Mel\'endez,  Santiago de Cali  760032,  Colombia}

\begin{abstract}
In this paper, we study the late time cosmology of a non-canonical scalar field ($k$-essence) coupled to a vector field in a Bianchi-I background. Specifically, we study three cases: canonical scalar field (quintessence) with exponential potential as a warm up, the dilatonic ghost condensate, and the Dirac Born Infeld field with exponentials throat and potential. By using a dynamical system approach, we show that anisotropic dark energy fixed points can be attractors for a suitable set of parameters of each model. We also numerically integrate the associated autonomous systems for particular initial conditions chosen in the deep radiation epoch. We find that the three models can give an account of an equation of state of dark energy close to $-1$ nowadays. However, a non-negligible spatial shear within the current observational bounds is possible only for the quintessence and the Dirac Born Infeld field. We also find that the equation of state of dark energy and the shear oscillate at late times, whenever the coupling of the $k$-essence field to the vector field is strong enough. The reason of these oscillations is explained in the appendix.
\end{abstract}


\maketitle

\section{Introduction} \label{Introduction}

It is an observational fact that the current Universe is expanding at an accelerated rate. This fact was firstly pointed out by type Ia supernovae (SNe Ia) surveys \cite{Riess:1998cb, Schmidt:1998ys, Perlmutter:1998np}, and later confirmed by several measurements of large-scale structures (LSS) \cite{Tegmark:2003ud, Tegmark:2006az}, cosmic microwave background (CMB) \cite{Spergel:2003cb, Aghanim:2018eyx}, and baryon acoustic oscillations (BAO)\cite{Percival:2007yw, Aubourg:2014yra}. Several projects have been able to characterize our observable Universe and now we know with good accuracy that it is highly homogeneous, isotropic and spatially flat at cosmological scales \cite{deBernardis:2000sbo, Jaffe:2003it, Aghanim:2018eyx}. The simplest model agreeing with the aforementioned observations is the standard $\Lambda$ Cold Dark Matter ($\Lambda$ CDM) model. In this model, the cosmological constant $\Lambda$ drives the Universe into its current accelerated expansion  \cite{Amendola:2015ksp, Bamba:2012cp}. Despite its success, it is plagued by theoretical and observational problems. One of the theoretical difficulties is the own nature of $\Lambda$, since if it is associated with the vacuum energy density of the Universe, the value predicted by the theory and the value obtained from observations differs by several tens orders of magnitude \cite{Weinberg:1988cp, Martin:2012bt}. This is known as the cosmological constant problem. On the observational side, the possible presence of the so-called CMB anomalies \cite{Perivolaropoulos:2014lua,Schwarz:2015cma}, firstly reported by WMAP \cite{Bennett:2010jb} and later by Planck \cite{Akrami:2019bkn, Akrami:2019izv}, suggest a departure from the isotropic $\Lambda$-CDM.\footnote{Wald's theorem: a cosmological constant can no support a prolonged anisotropic accelerated expansion \cite{Wald:1983ky}.} However, we would like to mention that since the statistical significance of these effects is small, the very existence of these anomalies is yet an open question \cite{Akrami:2019bkn, Schwarz:2015cma}. Moreover, it has been pointed out that the so called $H_0$ tension, i.e. the difference of the value obtained from SNe Ia and CMB measurements, could be alleviated by considering dynamical extensions to the cosmological standard model \cite{Perivolaropoulos:2021jda, DiValentino:2017iww, Guo:2018ans}.

The search for alternatives to the $\Lambda$ CDM model is generally split into two broad categories: modified gravity theories and dynamical dark energy. Recent observations, like the detection of gravitational waves \cite{TheLIGOScientific:2017qsa}, have put some models based on modifications to general relativity (GR) under serious observational pressure \cite{Creminelli:2017sry,Nersisyan:2018auj,Collett:2018gpf, Ezquiaga:2018btd, He:2018oai, Do:2019txf, Abbott:2018lct}. These modifications to gravity or dark energy fields are usually studied in a flat Friedmann-Lema\^itre-Robertson-Walker (FLRW) background, assuring a homogeneous and isotropic background evolution. However, the CMB anomalies (if they exist) imply a violation of the Universe's isotropy at large scales, meaning that a geometry different to that described by the FLRW metric should be considered \cite{Bennett:2010jb, Akrami:2019izv, Akrami:2019bkn}. It has been remarked that some of these CMB anomalies could be explained by the introduction of an anisotropic late-time accelerated expansion\footnote{We want to mention that some of these anomalies could be also the result of an unknown primordial mechanism acting during inflation \cite{Sanchez:2015wza}.} \cite{Battye:2009ze, Perivolaropoulos:2014lua, Schwarz:2015cma}. 

Among the proposals for anisotropic dark energy, we can find models that include  vector fields \cite{ArmendarizPicon:2004pm,Koivisto:2008ig, Koivisto:2008xf,Thorsrud:2012mu,Landim:2016dxh,Nakamura:2019phn,Gomez:2020sfz,GabrielGomez:2021mij}, p-forms \cite{Almeida:2019iqp, Guarnizo:2019mwf, Almeida:2018fwe}, non-Abelian gauge fields \cite{Mehrabi:2015lfa,Alvarez:2019ues,Orjuela-Quintana:2020klr, Guarnizo:2020pkj}, or inhomogeneous fields \cite{BuenoSanchez:2011wr, Perivolaropoulos:2014lua, Motoa-Manzano:2020mwe}. Although the most popular models of dark  energy are based on homogeneous scalar fields (quintessence) \cite{Ratra:1987rm, Caldwell:1997ii, Copeland:2006wr, Yoo:2012ug}, it is known that this kind of fields alone cannot source anisotropic stress, meaning that any initial spatial shear would be quickly damped and the Universe will isotropize. However, in Ref. \cite{Thorsrud:2012mu} was shown that if this quintessence is coupled to the canonical kinetic term of a vector field, anisotropic late time solutions are possible. This study was extended to the inflationary context in Ref. \cite{Ohashi:2013pca},  by considering the same coupling but to a non-canonical kinetic term for the scalar field (the so called $k$-inflation). In this case, it was shown that anisotropic inflationary solutions are a general property of the coupled $k$-inflation model. Here, inspired by Refs. \cite{Thorsrud:2012mu, Ohashi:2013pca}, we analyze the late time cosmology of a non-canonical scalar field coupled to the kinetic term of a vector field; i.e. we study the \emph{coupled $k$-essence model}. We find that, as in the inflationary case, anisotropic solutions are a generality of the model. In particular, we study three concrete models: the dilatonic ghost condensate (DGC), the Dirac Born Infeld (DBI) field with exponentials throat and potential, and the canonical scalar field with exponential potential for completeness; whose isotropic dynamics have been extensively studied in the literature (see Refs. \cite{Chiba:1999ka, Garousi:2000tr, ArmendarizPicon:2000ah, Malquarti:2003hn, ArkaniHamed:2003uy, Piazza:2004df, Bonvin:2006vc, Babichev:2007dw, Martin:2008xw, Guo:2008sz}, for example).

This paper is organized in the following way. In Sec. \ref{Model}, we present the action, the energy-momentum tensor, and the equations of motion for the fields. In Sec. \ref{Background}, the background equations for the Bianchi-I geometry are derived. Section \ref{DynamicalSystem} is dedicated to the dynamical system of the general coupled $k$-essence model. We study the canonical quintessence, the DGC, and the DBI models in Secs. \ref{Quintessence}, \ref{DGC Section}, and \ref{DBI Section}, respectively, performing the dynamical analysis and a numerical integration of the autonomous set for each one. Finally, our conclusions are presented in Sec. \ref{Conclusions}.

\section{General Model} \label{Model}

Consider the following action:
\begin{equation} \label{action}
S \equiv \int \text{d}^4\, x \sqrt{-g} \left( \mathcal{L}_\text{EH} + \mathcal{L} \right), 
\end{equation}
where $g$ is the determinant of the metric $g_{\mu\nu}$, \mbox{$\mathcal{L}_\text{EH} \equiv m_\text{P}^2 R / 2$} is the Einstein-Hilbert Lagrangian, $m_\text{P}$ is the reduced Planck mass, $R$ is the Ricci scalar, and $\mathcal{L}$ is the Lagrangian given by
\begin{equation} \label{Lagrangian}
\mathcal{L} \equiv - \frac{1}{4} f^2 (\phi) F_{\mu\nu} F^{\mu\nu} + \mathcal{P}(\phi, X) + \mathcal{L}_m + \mathcal{L}_r,
\end{equation}
where $F_{\mu\nu} \equiv \nabla_\mu A_\nu - \nabla_\nu A_\mu$ is the strength tensor associated with the vector field $A_\mu$, $\mathcal{P}$ is the Lagrangian of the $k$-essence scalar field $\phi$ and
\begin{equation}
X \equiv - \frac{1}{2} g^{\mu\nu} \nabla_\mu \phi \nabla_\nu \phi, 
\end{equation}
is the canonical kinetic term for this field. The function $f(\phi)$ is the coupling of the $k$-essence field to the vector field. $\mathcal{L}_m$ and $\mathcal{L}_r$ are the Lagrangians of matter and radiation fluids, respectively. This action was studied in Ref. \cite{Ohashi:2013pca} in the inflationary context (neglecting matter and radiation), where anisotropic solutions with constant shear were found.               

The energy-momentum tensor of the model is computed from $T_{\mu\nu} \equiv \mathcal{L} g_{\mu\nu} - 2 \frac{\delta \mathcal{L}}{\delta g^{\mu\nu}}$, giving
\begin{equation}
T_{\mu\nu} = T_{\mu\nu}^A + T_{\mu\nu}^\phi + T_{\mu\nu}^m + T_{\mu\nu}^r,
\end{equation}
where we have separated the contributions from the vector field and the scalar field as
\begin{align}
T_{\mu\nu}^A &\equiv f^2 (\phi) F_{\mu\rho} F_\nu^{\ \rho} - \frac{1}{4} g_{\mu\nu} f^2 (\phi) F_{\sigma\rho} F^{\sigma\rho}, \label{TA}\\
T_{\mu\nu}^\phi &\equiv \mathcal{P}_X \nabla_\mu \phi \nabla_\nu \phi + \mathcal{P} g_{\mu\nu}, \label{Tphi}
\end{align}
respectively,\footnote{We have used the shorthand notation $\mathcal{P}_X \equiv \frac{\partial \mathcal{P}}{\partial X}$.} while the energy-momentum tensor for the matter and radiation fluids are $T_{\mu\nu}^m$ and $T_{\mu\nu}^r$, respectively.

Varying the Lagrangian in Eq. \eqref{Lagrangian} with respect to $A_\nu$ we get the equation of motion for the vector field:
\begin{equation} \label{EoMA}
\nabla_\mu \left( f^2 (\phi)F^{\mu\nu} \right) = 0,
\end{equation}
and varying with respect to $\phi$ we get the equation of motion for the $k$-essence field as\footnote{Where we have denoted $\mathcal{P}_\phi \equiv \frac{\partial \mathcal{P}}{\partial \phi}$, $f_\phi \equiv \frac{\text{d} f}{\text{d} \phi}$.}
\begin{equation} \label{EoMphi}
\nabla_\mu \left( \mathcal{P}_X \nabla^\mu \phi \right) + \mathcal{P}_\phi - \left( 2 \frac{f_\phi}{f} \right) \frac{f^2(\phi)}{4} F_{\mu\nu} F^{\mu\nu} = 0.
\end{equation}
In the following section, we derive the dynamical equations of motion for the model in a specific anisotropic background.

\section{Background Equations of Motion} \label{Background}

The Lagrangian \eqref{Lagrangian} considers the interaction between a scalar field and a vector field. The introduction of such vector field violates rotational invariance.\footnote{We want to stress that a vector field \emph{does respect} isotropy if only the temporal component is considered \cite{DeFelice:2016yws, Heisenberg:2018mxx}.} Since the Universe is highly homogeneous and spatially flat \cite{Aghanim:2018eyx}, we assume an axially symmetric Bianchi-I spacetime, with a residual isotropy in the $(y, z)$ plane, such that the background geometry is given by
\begin{equation} \label{metric}
\text{d} s^2 = - \text{d} t^2 + a(t)^2 \left[ e^{-4 \sigma (t)} \text{d} x^2 + e^{2 \sigma (t)} \left( \text{d} y^2 + \text{d} z^2 \right) \right],
\end{equation}
where $a(t)$ is the average scale factor and $\sigma(t)$ is the geometrical shear, being both functions of the cosmic time $t$. A configuration for the vector field compatible with the symmetries of the metric is the following:
\begin{equation}\label{vconf}
A_\mu = \left( 0, A(t), 0, 0 \right),
\end{equation}
while the scalar field is only a function of time in order to preserve homogeneity, i.e. $\phi = \phi (t)$. With these choices, the kinetic term of the scalar field is $2 X = \dot{\phi}^2$ and the strength tensor of the vector field has only one independent component $F_{0 1} = \dot{A}$.

The energy density and pressure of the vector field ($\rho_A$ and $p_A$) and the $k$-essence field ($\rho_\phi$ and $p_\phi$) are obtained from Eqs. \eqref{TA} and \eqref{Tphi} as\footnote{Here, an overdot denotes a derivative with respect to the cosmic time $t$.}
\begin{align}
\rho_A &\equiv \frac{1}{2} f^2 (\phi) \frac{\dot{A}^2 e^{4\sigma}}{a^2}, \quad p_A \equiv \frac{1}{3} \rho_A, \\
\rho_\phi &\equiv 2 X \mathcal{P}_X - \mathcal{P}, \qquad \ p_\phi \equiv \mathcal{P}.
\end{align}

Using the gravitational field equations $m_\text{P}^2 G_{\mu\nu} = T_{\mu\nu}$, with $G_{\mu\nu}$ the Einstein tensor, the Friedman equations are
\begin{equation} \label{H2 eq}
3 m_\text{P}^2 H^2 = \rho_A + \rho_\phi + \rho_m + \rho_r + 3 m_\text{P}^2 \dot{\sigma}^2,
\end{equation}
\begin{equation} \label{Hdot eq}
- 2 m_\text{P}^2 \dot{H} = 2 X \mathcal{P}_X + \frac{4}{3} \rho_ A + \rho_m + \frac{4}{3} \rho_r + 6 m_\text{P}^2 \dot{\sigma}^2,
\end{equation}
where $H \equiv \dot{a} / a $ is the Hubble parameter. The evolution equation of the geometrical shear is obtained from the linear combination $m_\text{P}^2 \left( G^2_2 - G^1_1 \right) = T^2_2 - T^1_1$ as
\begin{equation}
3 m_\text{P}^2 \ddot{\sigma} + 9 m_\text{P}^2 H \dot{\sigma} = 2 \rho_A, 
\end{equation}
We can identify that the matter source for the shear is the energy density of the vector field. Although this source is different from zero even in the case when there is no coupling to the scalar field; i.e. when $f(\phi) = 1$, we will see later that it is precisely this interaction which allows late times anisotropic solutions with interesting oscillatory behaviors.

Now, replacing the metric \eqref{metric} and the configurations of the fields in the equation \eqref{EoMphi}, we get
\begin{align}
0 &= \left( 2 X \mathcal{P}_{XX} + \mathcal{P}_X \right) \ddot{\phi} + 3 H \mathcal{P}_X \dot{\phi} + \mathcal{P}_{X\phi} \dot{\phi}^2 \nonumber \\
 &- \mathcal{P}_\phi - \left(2 \frac{f_\phi}{f} \right) \frac{f^2(\phi)}{2} \frac{\dot{A}^2 e^{4\sigma}}{a^2},
\end{align}
and from Eq. \eqref{EoMA} we get that the speed of the vector field is given by
\begin{equation}
\dot{A} = c_A \frac{e^{-4\sigma}}{a f^2(\phi)},
\end{equation}
with $c_A$ a constant.
It is known that isotropic $k$-essence models allow for scaling solutions whenever the Lagrangian has the form
\begin{equation}
\mathcal{P}(\phi, X) \equiv X G(Y), \quad Y \equiv X e^{\lambda \phi / m_\text{P}},
\end{equation}
where $\lambda$ is a constant \cite{Amendola:2015ksp}. Here, we assume this form of the Lagrangian, and then the equation of motion of the scalar field is simplified as
\begin{equation}
\ddot{\phi} + 3 H Q \mathcal{P}_X \dot{\phi} + Q (\mathcal{P}_{X\phi} - \mathcal{P}_\phi) - 2 Q \frac{f_\phi}{f} \rho_A = 0,
\end{equation}
where $Q(Y) \equiv (2 X \mathcal{P}_{XX} + \mathcal{P}_X)^{-1}$.
Following \cite{Ohashi:2013pca}, we define
\begin{equation}
G_n \equiv Y^n \frac{\text{d}^n G}{\text{d} Y^n},
\end{equation}
in order to separate the terms which are function only of $Y$. For the derivatives of the Lagrangian $\mathcal{P}$ we obtain
\begin{align}
\mathcal{P}_\phi &= \frac{\lambda}{m_\text{P}} X G_1 (Y), \nonumber \\
\mathcal{P}_X &= G(Y) + G_1(Y), \nonumber \\ 
\mathcal{P}_{X\phi} &= \frac{\lambda}{m_\text{P}} \left[ 2G_1(Y) + G_2(Y) \right], \nonumber \\
X\mathcal{P}_{XX} &= 2 G_1(Y) + G_2(Y), \nonumber \\
Q(Y) &= \left[ G(Y) + 5 G_1(Y) + 2 G_2(Y) \right]^{-1},
\end{align}
and thus the equation of motion for the scalar field can be rewritten as
\begin{align} \label{EoM phi}
0 &= \ddot{\phi} + 3 H Q(Y) \mathcal{P}_X (Y) \dot{\phi} - 2 Q(Y) \left( \frac{f_\phi}{f} \right) \rho_A \nonumber \\
 &+ \frac{\lambda}{m_\text{P}} X \left[ 1 - (G(Y) + 2 G_1(Y)) Q(Y) \right].
\end{align}

\section{Dynamical system} \label{DynamicalSystem}

Our next task will be to recast the background equations in an autonomous system and find their fixed points, in order to know the asymptotic behavior of the model \cite{Wainwright2009}. Therefore, we introduce the following dimensionless variables:
\begin{equation*}
x \equiv \frac{\dot{\phi}}{\sqrt{6} m_\text{P} H}, \quad z^2 \equiv \frac{\rho_A}{3 m_\text{P}^2 H^2}, \quad \Omega_m \equiv \frac{\rho_m}{3 m_\text{P}^2 H^2},
\end{equation*}
\begin{equation} \label{variables}
\Omega_r \equiv \frac{\rho_r}{3 m_\text{P}^2 H^2}, \quad \Sigma \equiv \frac{\dot{\sigma}}{H},
\end{equation}
such that the first Friedman equation \eqref{H2 eq} becomes the constraint
\begin{equation} \label{Constraint}
1 = x^2 \left[ 2 \mathcal{P}_X(Y) - G(Y) \right] + z^2 + \Omega_m + \Omega_r + \Sigma^2,
\end{equation}
from which, we can write $\Omega_m$ in terms of the other variables. From the second Friedman equation in Eq. \eqref{Hdot eq}, we compute the deceleration parameter \mbox{$q \equiv - 1 - \dot{H}/H^2$}, obtaining
\begin{equation} \label{deceleration}
q = \frac{1}{2} \left[ 1 + 3 x^2 G(Y) + z^2 + \Omega_r + 3 \Sigma^2 \right].
\end{equation}
For the dark sector, we define the density and pressure of dark energy as
\begin{align}
\rho_\text{DE} &\equiv \rho_A + \rho_\phi + 3 m_\text{P}^2 \dot{\sigma}^2, \\ 
p_\text{DE} &\equiv p_A + p_\phi + 3 m_\text{P}^2 \dot{\sigma}^2,
\end{align}
and thus the equation of state of dark energy, \mbox{$w_\text{DE} \equiv p_\text{DE} / \rho_\text{DE}$}, is given by
\begin{equation}\label{DEeqs}
w_\text{DE} = - 1 + \frac{2}{3} \frac{3 x^2 \mathcal{P}_X + 2 z^2 + 3 \Sigma^2}{x^2( \mathcal{P}_X + G_1) + z^2 + \Sigma^2}.
\end{equation}
Note that with this choice of including the geometrical shear in $\rho_\text{DE}$ and $p_\text{DE}$ we can write the continuity equation of dark energy as
\begin{equation}
\dot{\rho}_\text{DE} + 3 H \left( \rho_\text{DE} + p_\text{DE} \right) = 0,
\end{equation}
which is the usual form. In the case that this anisotropic contribution is not consider in the definition of the density and pressure of dark energy, we would obtain an equation of the form \mbox{$\dot{\rho}_\text{DE} + 3 H \left(\rho_\text{DE} + p_\text{DE} \right) \propto \dot{g}_{ij}\Pi^{ij}$,} where $\Pi^{ij}$ is the trace-free part of the energy-momentum tensor.

\subsection{Autonomous System}

We want to write the autonomous set of equations for the dynamical variables in Eq. \eqref{variables}. Therefore, it is necessary to choose a specific coupling. To ease the computations we consider an exponential function of the form
\begin{equation}
f(\phi) = f_0 e^{-\mu \phi / m_\text{P}},
\end{equation}
where $\mu$ is a constant. By taking the derivative of the dimensionless variables with respect to the number of $e$-folds \mbox{$\text{d} N \equiv H \text{d} t$,} we get the following autonomous system:\footnote{Here, a prime denotes a derivative with respect to $N$.}
\begin{align}
x' &= x(q + 1) - \frac{\sqrt{6}}{2} \lambda x^2 \nonumber \label{x eq}\\
 &- \frac{\sqrt{6}}{2} Q \left[ \sqrt{6} x \mathcal{P}_X - \lambda x^2 \left( \mathcal{P}_X + G_1 \right) + 2 \mu z^2 \right], \\
z' &= z(q - 1) + \sqrt{6} \mu x z - 2 z \Sigma, \\
\Sigma' &= \Sigma (q - 2) + 2 z^2, \label{Sigma eq} \\
\Omega_r' &= 2 \Omega_r (q - 1). \label{r eq}
\end{align}
We note that this system of differential equations is not closed since the function $Y$ is not written in terms of the variables. For this function, as we will see in the next sections, it is necessary to choose a specific variable depending on the form of the Lagrangian in order to avoid possible singularities in the system, which are related to the term $x^2 \left[ 2 \mathcal{P}_X(Y) - G(Y) \right]$ in Eq. (\ref{Constraint}).

\section{Anisotropic Quintessence} \label{Quintessence}

As a warm up, we study firstly the coupled canonical quintessence with an exponential potential. In this case, the Lagrangian is
\begin{equation}
\mathcal{P} (\phi, X) \equiv X - c \, m_\text{P}^4 e^{-\lambda \phi / m_\text{P}},
\end{equation}
with $c$ a constant. We have that $\mathcal{P}_X = 1$, while the function characterizing the Lagrangian is
\begin{equation}
G(Y) = 1 - c \frac{m_\text{P}^4}{Y}.
\end{equation}
From the later we get $G_1(Y) = 1 - G(Y)$ and \mbox{$Q(Y) = 1$.} This model was studied in \cite{Thorsrud:2012mu}, where a coupling between CDM and the scalar field was also considered, which we neglect here.

We note that the term $x^2(2 \mathcal{P}_X - G)$ in the Friedman constraint \eqref{Constraint} can yield singularities depending on the variables we choose to define $Y$. In this case, we can use the dimensionless variable\footnote{We want to stress that avoiding singularities in the Friedman constraint by choosing a suitable variable is extremely important, since some fixed points could be ``hidden'' in these divergences \cite{Boehmer:2008av}.}
\begin{equation}
y \equiv c_y \frac{m_\text{P} \, e^{- \lambda \phi / 2 m_\text{P}}}{\sqrt{3} H},
\end{equation}
where $c_y \equiv c^{1/2}$, such that
\begin{equation}
x^2 \left( 2 \mathcal{P}_X - G \right) = x^2 + y^2.
\end{equation}
The evolution equation of this new variable is
\begin{equation} \label{y eq}
y' = y (q + 1) - \frac{\sqrt{6}}{2} \lambda x y.
\end{equation}
The fixed points of the autonomous set are obtained by setting $x' = 0$, $y' = 0$, $z' = 0$, $\Omega_r' = 0$, and $\Sigma' = 0$, in Eqs. \eqref{x eq}-\eqref{r eq} and \eqref{y eq}. Then, the stability of these points can be known by perturbing the autonomous set around them. Up to linear order, the perturbations $\delta \mathcal{X} = ( \delta x, \delta y, \delta z, \delta \Sigma , \delta \Omega_r)$ satisfy the differential equation,
\begin{equation}
\delta \mathcal{X}' = \mathbb{M} \, \delta \mathcal{X},
\end{equation}
where $\mathbb{M}$ is a $5 \times 5$ Jacobian matrix. The sign of the real part of the eigenvalues $\eta_{1, 2, 3, 4, 5}$ of $\mathbb{M}$ determines the stability of the
point. A fixed point is an attractor, or sink, if the real part of all the eigenvalues are negative. If there is a mix of positive and negative eigenvalues in a point, that point is called a saddle. If the real part of all the eigenvalues are positive the fixed point is called a repeller or source.
\begin{center}
\begin{table*}
\begin{centering}
\begin{tabular}{cccccccc}
\toprule 
Fixed Point & $x$ & $y$ & $z$ & $\Omega_{r}$ & $\Omega_{m}$ & $\Sigma$ & $q$ \tabularnewline
\midrule
\midrule 
(\emph{R-1}) & 0 & 0 & 0 & 1 & 0 & 0 & 1 \tabularnewline
\midrule
(\emph{R-2}) & $\frac{2 \sqrt{6}}{3 \lambda}$ & $\frac{2 \sqrt{3}}{3 \lambda}$ & 0 & $1 - \frac{4}{\lambda^2}$ & 0 & 0 & 1 \tabularnewline
\midrule
(\emph{R-3}) & $\frac{2 \sqrt{6}}{3 \lambda}$ & $\frac{\sqrt{4 + 6\mu^2}}{\sqrt{3}\lambda}$ & $\sqrt{\frac{\mu}{\lambda}}$ & $1 - \frac{4 + \lambda \mu + 6 \mu^2}{\lambda^2}$  & 0 & $2 \frac{\mu}{\lambda}$ & 1 \tabularnewline
\midrule
(\emph{M-1}) & 0 & 0 & 0 & 0 & 1 & 0 & 1/2 \tabularnewline
\midrule
(\emph{M-2}) & $\frac{\sqrt{6}}{2 \lambda}$ & $\frac{\sqrt{6}}{2 \lambda}$ & 0 & 0 & $1 - \frac{3}{\lambda^2}$ & 0 & 1/2 \tabularnewline
\midrule
(\emph{M-3}) & $\frac{\sqrt{6}}{2 \lambda}$ & $\frac{\sqrt{12 - 3 \lambda \mu + 18 \mu^2}}{2\sqrt{2}\lambda}$ & $\sqrt{\frac{9 \mu}{8\lambda} - \frac{3}{16}}$ & 0 & $\frac{3( -8 + 3\lambda^2 - 12 \mu^2)}{8 \lambda^2}$ & $\frac{6 \mu - \lambda}{4 \lambda}$ & 1/2 \tabularnewline
\midrule
\bottomrule
\end{tabular}
\par\end{centering}
\caption{Fixed points for the canonical scalar field. The points are labelled according to the cosmological regime as (\emph{R- }) (radiation) and (\emph{M- }) (matter).}
\label{Table Q}
\end{table*}
\par\end{center}
In general, we would discuss the fixed points relevant to the radiation ($\Omega_r \approx 1$, $w_\text{eff} \approx 1/3$), matter ($\Omega_m \approx 1$, $w_\text{eff} \approx 0$), and dark energy eras ($\Omega_\text{DE} \approx 1$, $w_\text{eff} < -1/3$), where $w_\text{eff}$ is the effective equation of state given by
\begin{equation} \label{weff}
w_\text{eff} = - 1 - \frac{2}{3} \frac{\dot{H}}{H^2} \frac{1}{1 - \Sigma^2} - 2 \frac{\Sigma^2}{1 - \Sigma^2},
\end{equation}
which can be related to the deceleration parameter\footnote{Since CMB observations imply $\Sigma \ll 1$, then $w_\text{eff} \approx (2q - 1)/3$ and  we have the following easy identifications: $q = 1 \Rightarrow w_\text{eff} \approx 1/3$, $q = 1/2 \Rightarrow w_\text{eff} \approx 0$, and $q < 0 \Rightarrow w_\text{eff} < -1/3$, for radiation, matter, and dark energy dominated epochs respectively.} by noting that $- \dot{H} / H^2 = 1 + q$. However, since the full phase space of this model was deeply explored in Ref. \cite{Thorsrud:2012mu}, and since we are mainly interested in the late cosmology of the model, we only analyze the accelerated solutions while briefly mentioning the other points. For more details about the viable cosmological trajectories of the model, please see Ref. \cite{Thorsrud:2012mu}. In what follows, we refer to each point by its name, which is defined as (\emph{R-n}), (\emph{M-n}) or (\emph{DE-n}) $-$ depending on whether it corresponds to a radiation, matter or dark energy dominated universe $-$ followed by a number $n$. The radiation and matter points are gathered in Table \ref{Table Q}.

\subsection{Dark Energy Fixed Points}

\begin{itemize}
\item \textbf{(\emph{DE-1}): Isotropic Dark Energy}
\end{itemize}
This is the usual isotropic dark energy fixed point where the variables take the values
\begin{equation}
x = \frac{\lambda}{\sqrt{6}}, \ y = \sqrt{1 - \frac{\lambda^2}{6}}, \ z = 0, \ \Sigma = 0, \ \Omega_r = 0,
\end{equation}
with $\Omega_m = 0$. We can see that the density of the vector vanishes, therefore the shear is zero. The effective equation of state is 
\begin{equation}
w_{\text{eff}} = w_\text{DE} = - 1 + \lambda^2 / 3
\end{equation}
meaning that this point is an accelerated solution for $\lambda^2 < 2$, as expected. The stability of the point, however, change with the introduction of the \mbox{parameter $\mu$.} The eigenvalues of $\mathbb{M}$ evaluated at the point are
\begin{align}
\eta_{1, 2} &= - 3 + \frac{\lambda^2}{2}, \quad \eta_3 = - 4 + \lambda^2, \nonumber \\ 
\quad \eta_4 &= - 3 + \lambda^2, \quad \eta_5 = - 2 + \frac{\lambda^2}{2} + \lambda \mu.
\end{align}
For $\lambda \geq 0$ these eigenvalues are all negative whenever\footnote{The symbol $\land$ stands for the logic ``AND''.}
\begin{equation} \label{QDE1 conditions}
\lambda^2 < 2 \quad \land \quad \mu \leq \frac{4 - \lambda^2}{2\lambda}.
\end{equation}
Note that $\lambda \approx 0$ in order to have $w_\text{DE} \approx -1$, as required by observations \cite{Aghanim:2018eyx}. We can estimate the bound for $\mu$ from this information. For instance, choosing $\lambda = 0.1$, we get $w_\text{DE} \approx -0.997$ and $\mu \leq 19.95$. Under these exemplifying conditions, this point will be an attractor of the system.

\begin{itemize}
\item \textbf{(\emph{DE-2}): Anisotropic Dark Energy}
\end{itemize}

\begin{figure}[t!]
\includegraphics[width=0.92\linewidth]{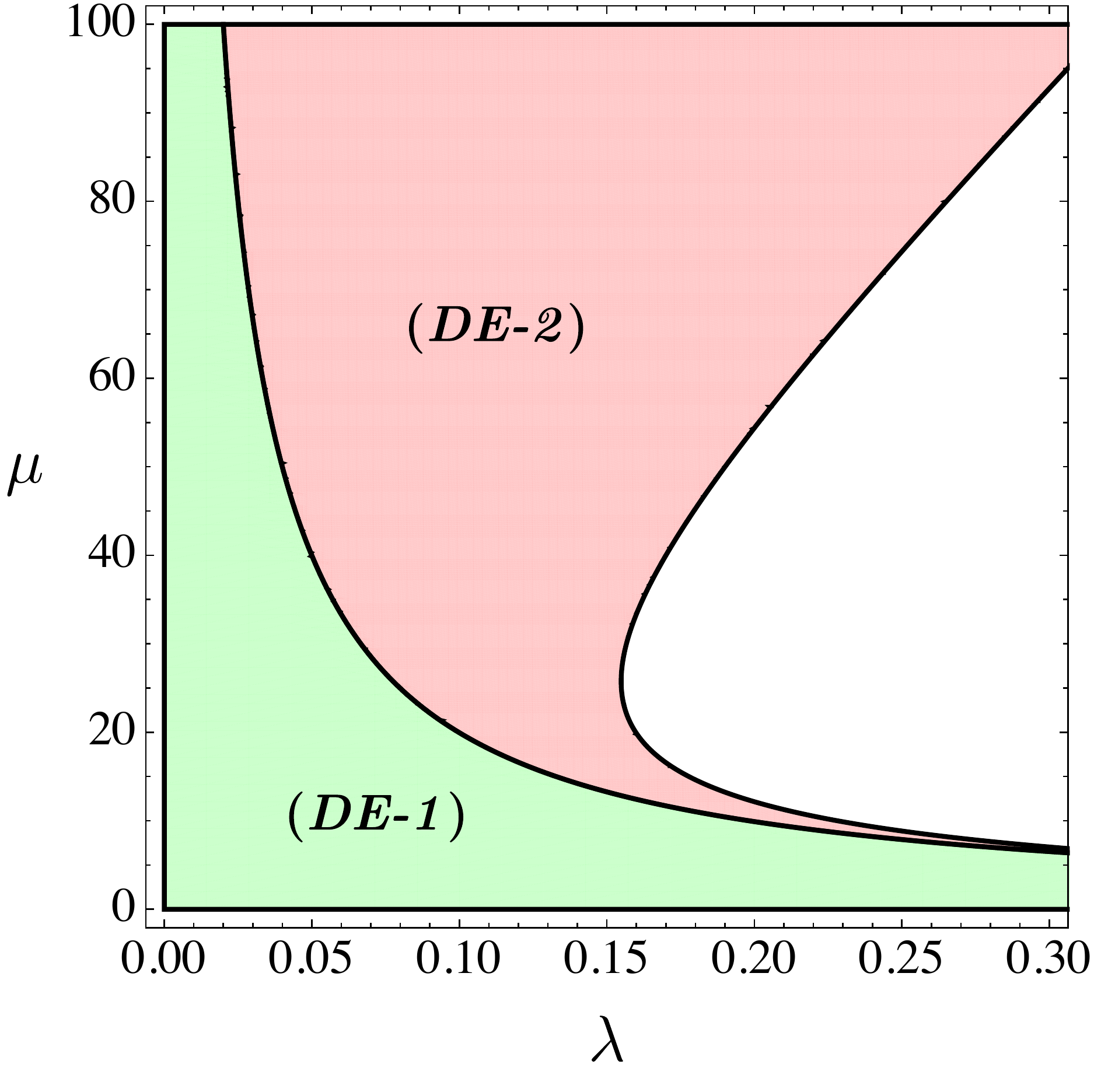}    
\caption{Stability regions for the dark energy dominated points (\emph{DE-1}) and (\emph{DE-2}). Each color represents a ($\lambda, \mu$) parameter region where the indicated fixed point is an attractor with $-1 \leq w_\text{DE} \leq -0.95$ and the shear is small $\Sigma \leq 10^{-3}$. These regions are separated by the bifurcation curve \mbox{$2\lambda \mu = 4 - \lambda^2$}.}
\label{Regions}
\end{figure}

The values of the variables are
\begin{align} \label{Q DE2 variables}
x &= \frac{2 \sqrt{6} \left( \lambda + 2 \mu \right)}{8 + (\lambda + 2\mu)(\lambda + 6\mu)}, \nonumber \\ 
y &= \frac{\sqrt{6(8 - (\lambda - 6\mu)(\lambda + 2\mu))(2 + \mu (\lambda + 2 \mu))}}{8 + (\lambda + 2\mu)(\lambda + 6\mu)}, \nonumber \\
z &= \frac{\sqrt{3(8 - (\lambda - 6\mu)(\lambda + 2\mu))(\lambda^2 + 2 \mu \lambda - 4)}}{8 + (\lambda + 2\mu)(\lambda + 6\mu)}, \nonumber \\
\Sigma &= 2 \frac{\lambda^2 + 2 \mu \lambda - 4}{8 + (\lambda + 2\mu)(\lambda + 6\mu)},
\end{align}
with $\Omega_r = 0$, $\Omega_m = 0$, and
\begin{equation} \label{wDEQ}
w_\text{DE} = - 1 + \frac{4\lambda (\lambda + 2 \mu)}{8 + (\lambda + 2\mu)(\lambda + 6\mu)}.
\end{equation}
This point is a viable solution if the variables take real values, i.e. $y^2, z^2 > 0$, and it is an accelerated solution if $w_\text{DE} < - 1/3$. These conditions allow us to determine the parameter space $(\lambda, \mu)$ where the point is a cosmological solution. In this case, they imply
\begin{equation} \label{QDE2}
\lambda < \sqrt{2} \quad \land \quad \mu \geq \frac{4 - \lambda^2}{2 \lambda}.
\end{equation}
We see that the condition on $\mu$, in this case, is exactly the opposite of the condition on $\mu$ for the isotropic point in \mbox{Eq. \eqref{QDE1 conditions}.} This means that when the anisotropic point (\emph{DE-2}) is an accelerated solution, the isotropic point (\emph{DE-1}) is a saddle. Also note that, since an equation of state of dark energy near to $-1$ requires\footnote{There is another possibility, namely, $\lambda \sim - 2\mu$, but we do not pursue this rather fine-tuned case.} $\lambda \sim 0$, then this point needs a large $\mu$. This can be clarified by noting that a small shear, as also required by observations \cite{Campanelli:2010zx, Amirhashchi:2018nxl}, can be satisfied if $\mu \gg \lambda $ and $\mu \gg 1$. Assuming the later, the shear and the equation of state of dark energy can be approximated to first order by
\begin{equation}
\Sigma \approx \frac{\lambda}{3 \mu} - \frac{2}{3 \mu^2}, \quad w_\text{DE} \approx - 1 + \frac{2 \lambda}{3 \mu},
\end{equation}
and the eigenvalues of $\mathbb{M}$ are approximated by
\begin{align}
\eta_1 &\approx - 3 + \frac{\lambda}{\mu}, \quad \eta_2 \approx - 3 + 2 \frac{\lambda}{\mu}, \nonumber \\
\eta_3 &\approx - 4 + 2 \frac{\lambda}{\mu}, \quad \eta_{4, 5} \approx - \frac{3}{2} + \frac{\lambda}{2 \mu}.
\end{align}
Therefore, for small $\lambda$ and \emph{large} $\mu$ this point is an accelerated solution with $w_\text{DE} \approx -1$, a small shear, and it is an attractor, confirming what we said above. In Fig. \ref{Regions}, we can see that the two possible dark energy points are separated by a bifurcation curve.

\subsection{Matter Fixed Points}

The point (\emph{M-1}) is the usual saddle isotropic point with no contribution of radiation or dark energy, while (\emph{M-2}) is the usual isotropic matter-dark energy scaling, where the CMB bound on early dark energy \cite{Aghanim:2018eyx} $\Omega_\text{DE} < 0.02$ around the redshift $z_r = 50$ requires $\lambda > 12. 25$. The point (\emph{M-3}) is a scaling solution with non-negligible shear. However, a small shear requires $\lambda \approx 6 \mu$, which is a kind of fine-tuning, and the CMB bound gives the same bound for $\lambda$ as in (\emph{M-2}). Since the existence and stability of the dark energy points require a small $\lambda$, the points (\emph{M-2}) and (\emph{M-3}) are ruled out as possible matter dominated points, and thus the isotropic point (\emph{M-1}) will be dynamically selected.

\subsection{Radiation Fixed Points}

The point (\emph{R-1}) is the usual saddle isotropic point with no contribution of matter or dark energy, while (\emph{R-2}) is the usual isotropic radiation-dark energy scaling where the big-bang nucleosynthesis (BBN) bound on early dark energy \cite{Bean:2001wt} $\Omega_\text{DE} < 0.045$ requires $\lambda > 9.43$. The point (\emph{R-3}) is a scaling solution with non-negligible shear. The BBN bound puts the same bound for $\lambda$ as in (\emph{R-2}) and a small shear requires a very small $\mu$. Since the existence and stability of the dark energy points require a small $\lambda$, the points (\emph{R-2}) and (\emph{R-3}) are ruled out as possible radiation dominated points, and thus the isotropic point (\emph{R-1}) will be dynamically selected.

The later analysis shows us that the cosmological trajectory of this model will be

\begin{center}
(\emph{R-1}) \ $\rightarrow$ \ (\emph{M-1}) \ $\rightarrow$ \ (\emph{DE-1})/(\emph{DE-2}).
\end{center}

\subsection{Cosmological evolution} \label{CE}

In this section we solve numerically the autonomous set in Eqs. \eqref{x eq}-\eqref{r eq} and \eqref{y eq}. We assume that, prior to the radiation epoch, the Universe underwent an inflationary period which perfectly smoothed any initial spatial shear or inhomogeneities. Therefore, we choose $\Sigma_i = 0$ as an initial condition, such that the starting point for any cosmological trajectory is from the isotropic radiation point (\emph{R-1}). We will choose parameters $\lambda$ and $\mu$ such that the attractor point is given by (\emph{DE-2}). In particular, we have chosen $\lambda = 0.1$, and $\mu = 10^3$ from the bound in Eq. \eqref{QDE2}. Having this in mind, we set
\begin{equation*}
x_i = 10^{-15}, \quad y_i = 2 \times 10^{-14}, \quad z_i = 10^{-12},
\end{equation*}
\begin{equation} \label{Q init cond}
\Sigma_i = 0, \quad \Omega_{r_i} = 0.99995,
\end{equation}
as initial conditions at redshift $z_r = 6.57 \times 10^7$, and we have integrated the system up to $z_r \rightarrow -1$.

\begin{figure}[t!]
\includegraphics[width=0.95\linewidth]{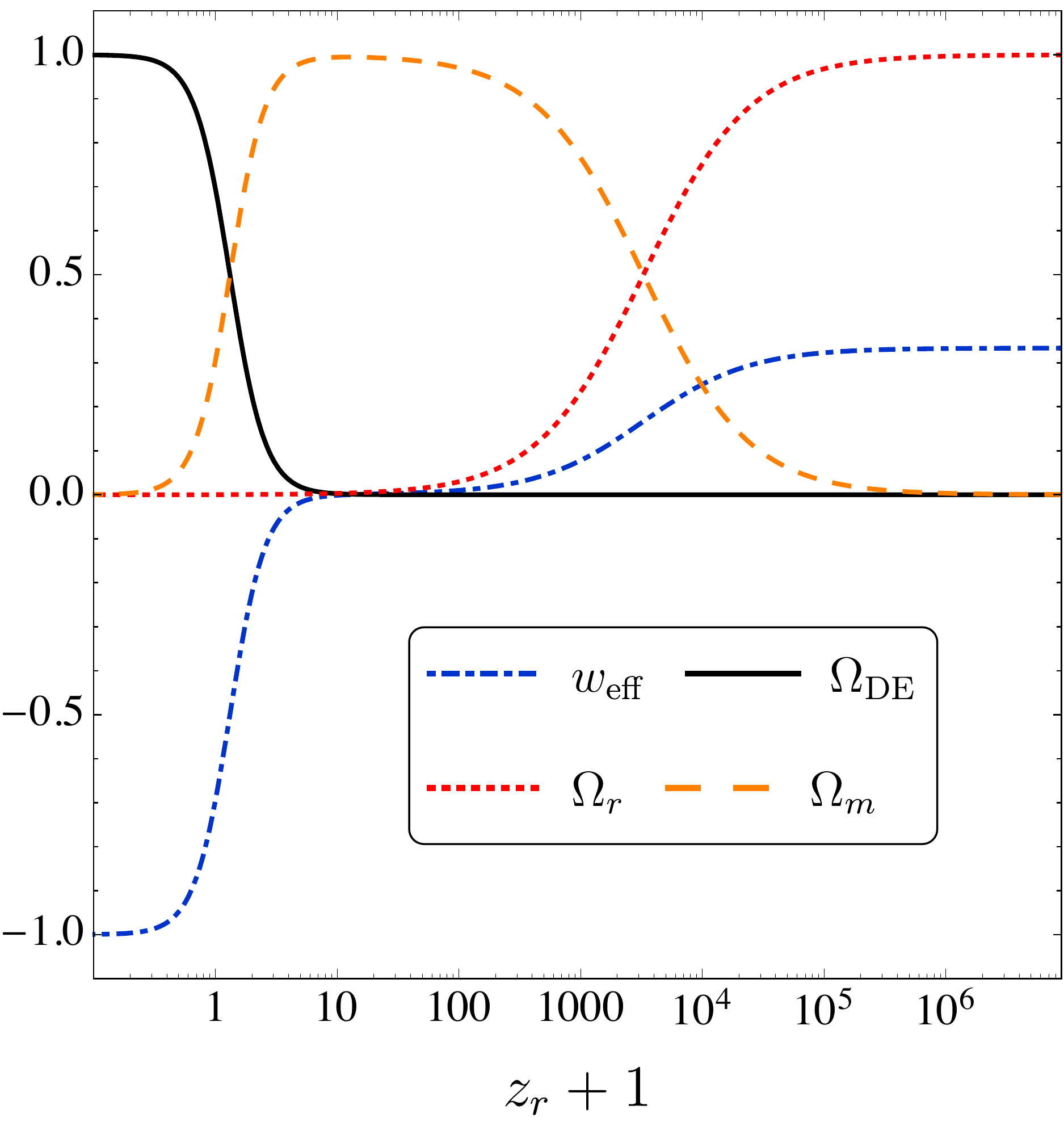}    
\caption{Evolution of the density parameters, the effective equation of state, and the equation of state of dark energy during the whole expansion history. The initial conditions were chosen deep in the radiation era at the redshift $z_r = 6.57 \times 10^7$. The Universe passes through radiation dominance at early times (red dotted line), followed by a matter dominance (light brown dashed line), and ends in the dark energy dominance (black solid line) characterized by $w_\text{eff} \simeq -1$ (blued dot-dashed line).}
\label{QAbundances}
\end{figure}

In Fig. \ref{QAbundances}, we plot the cosmological evolution of the density parameters of radiation, matter, and dark energy, and also the effective equation of state given in Eq. \eqref{weff}. In particular, Fig. \ref{QAbundances} shows that the radiation-dominated epoch ($\Omega_r \approx 1$ and $w_\text{eff} \approx 1/3$) runs from $z_r \gsim 10^6$ to $z_r \approx 3200$ where the radiation–matter transition occurs. Moreover, $\Omega_\text{DE} \approx 3.52 \times 10^{-11}$ during this transition, obeying the BBN constraint $\Omega_\text{DE} < 0.045$. The length of this radiation phase is in agreement with the constraint given in Ref. \cite{Alvarez:2019ues}. From $z_r \approx 3200$, the Universe is dust-dominated ($\Omega_m \approx 1$ and $w_\text{eff} \approx  0$) until $z_r \approx 0.3$ at the matter-dark energy transition. The contribution of the dark sector is $\Omega_\text{DE} \approx 1.58 \times 10^{-5}$ at $z_r = 50$, value which is within the CMB bound $\Omega_\text{DE} < 0.02$. The dark energy-dominance ($\Omega_\text{DE} \approx 1$ and $w_\text{eff} < -1/3$) starts from $z_r = 0.3$ and on into the future, agreeing with the results given by the dynamical system analysis, i.e. (\emph{DE-2}) is an attractor. This is further supported by the fact that the values of $x$, $y$, $z$, $\Sigma$ and $w_\text{DE}$ are those predicted by the dynamical system. Explicitly, $x \approx 8.2 \times 10^{-4}$, $y \approx 0.99998$, $z \approx 7 \times 10^{-3}$, $\Sigma \approx 3.27 \times 10^{-5}$, and $w_\text{DE} \approx -0.999933$ in the far future ($z_r \rightarrow -1$), which are consistent with the values computed from the (\emph{DE-2}) in Eqs. \eqref{Q DE2 variables} and \eqref{wDEQ}. In summary, the cosmological trajectory of the model is

\begin{center}
(\emph{R-1}) \ $\rightarrow$ \ (\emph{M-1}) \ $\rightarrow$ \ (\emph{DE-1}) \ $\rightarrow$ \ (\emph{DE-2}).
\end{center}

\begin{figure}[t!]
\includegraphics[width=0.92\linewidth]{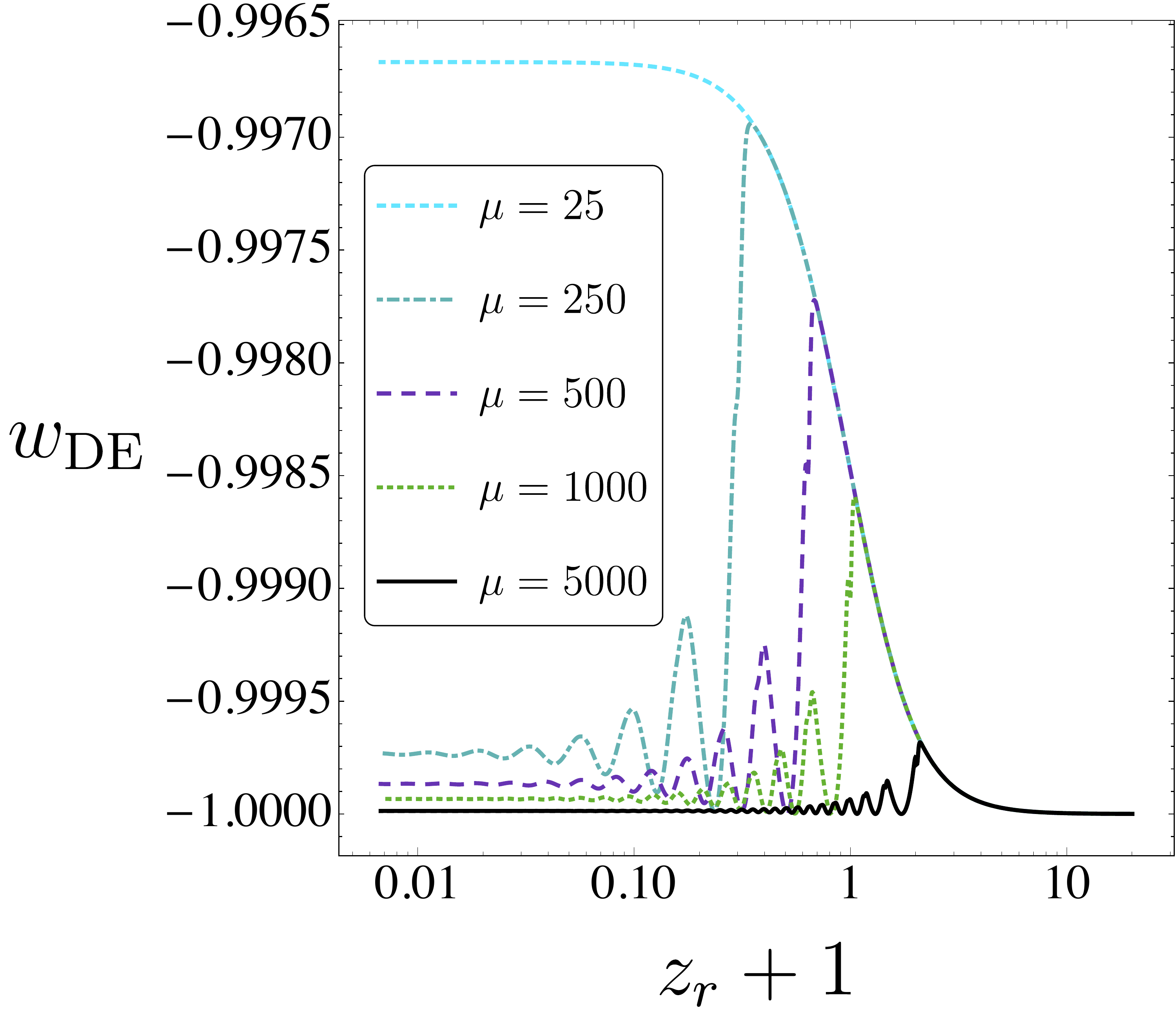}    
\caption{Evolution of the equation of state of dark energy $w_\text{DE}$ around $z_r = 0$ for different values of the parameter $\mu$, while $\lambda$ is fixed and the initial conditions are the same given in Eq. \eqref{Q init cond}. }
\label{QOscwDE}
\end{figure}

\begin{figure}[t!]
\includegraphics[width=0.92\linewidth]{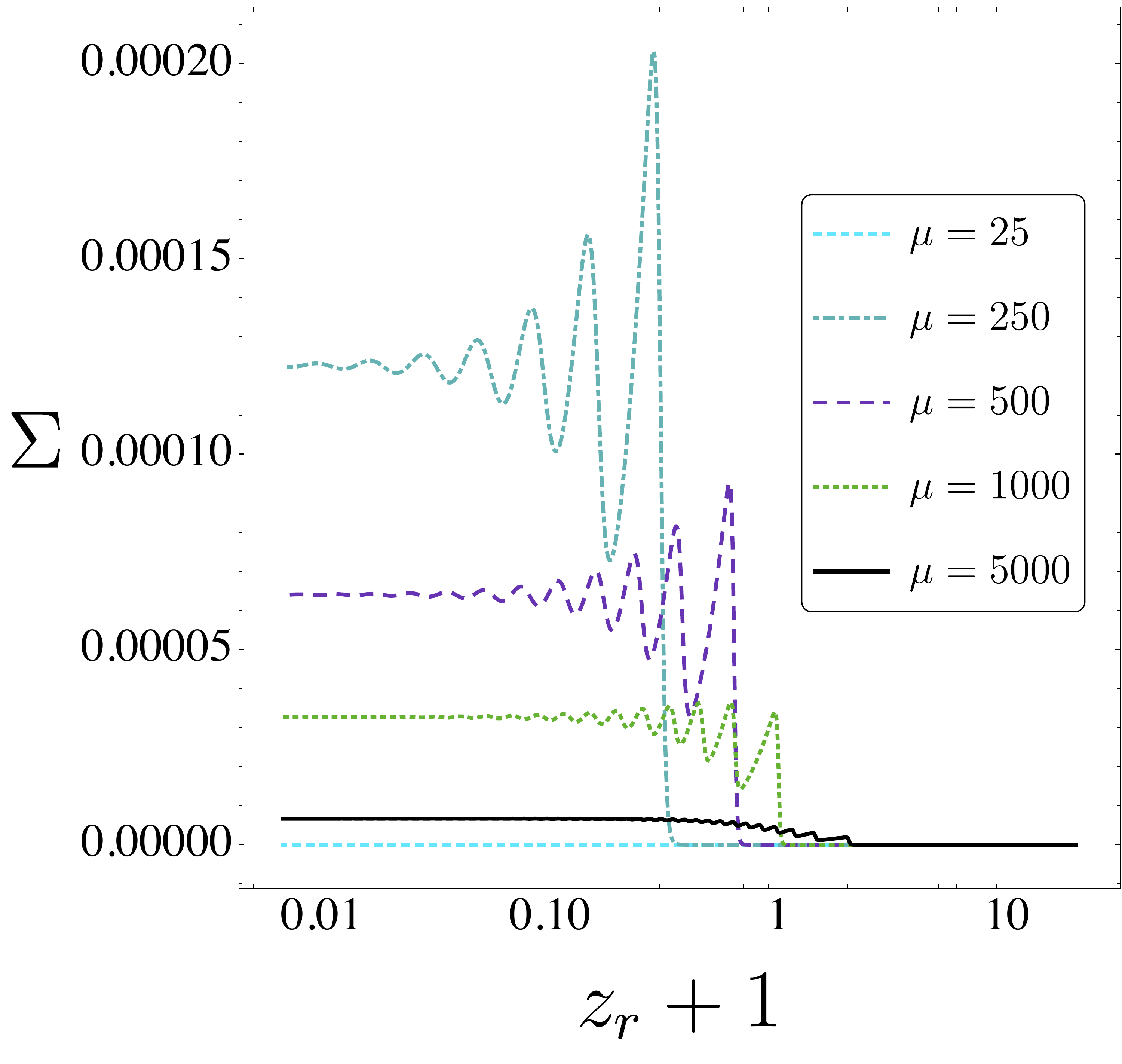}    
\caption{Evolution of the shear $\Sigma$ around $z_r = 0$ for different values of the parameter $\mu$, while $\lambda$ is fixed and the initial conditions are the same given in Eq. \eqref{Q init cond}.}
\label{QOscShear}
\end{figure}

In Figs. \ref{QOscwDE} and \ref{QOscShear}, we show the late-time evolution of the equation of state of dark energy and the shear for a fixed $\lambda = 0.1$ and varying $\mu$. We can see that the equation of state of dark energy is very close to $-1$ and that $\Sigma$ is within the observational bound today, namely, $|\Sigma_0| \leq 10^{-3}$ \cite{Campanelli:2010zx, Amirhashchi:2018nxl}. We also see an interesting behavior of the equation of state and the shear: they oscillate. This can be understood as follows. Note that when $\mu = 25$, there is no oscillations in $w_\text{DE}$ since in this case the system spend more time in the saddle isotropic point (\emph{DE-1}). For larger values of $\mu$ the coupling of the vector field to the scalar field is stronger, and the contribution of the vector field to the density is greater. It can be shown that at some late-time, the density of the vector field is comparable to the kinetic energy of the scalar field; i.e. $z \approx x$, and hence the scalar field obeys the equation
\begin{equation}
\ddot{\phi}(t) + \gamma \dot{\phi}(t) + \omega_0^2 \phi(t) = F_0,
\end{equation}
which is the equation of a constant forced and damped harmonic oscillator with constants $\gamma$, $\omega_0$, and $F_0$ (see Appendix \ref{App} for details). This oscillatory behavior of the scalar field, and the density of the vector field, is the reason of the oscillations of $\Sigma$ and $w_\text{DE}$.

\section{Anisotropic Dilatonic Ghost Condensate} \label{DGC Section}

Next, we study the dilatonic ghost condensate (DGC) model whose Lagrangian reads
\begin{equation}
\mathcal{P} (\phi, X) \equiv - X + b \, e^{\lambda \phi / m_\text{P}} \frac{X^2}{m_\text{P}^4 },
\end{equation}
and thus we have $\mathcal{P}_X(Y) = - 1 + 2 b \, Y/m_\text{P}^4$, while the function characterizing the Lagrangian is
\begin{equation}
G(Y) = - 1 + b \frac{Y}{m_\text{P}^4},
\end{equation}
where $b$ is a constant. From the later we get $G_1(Y) = 1 + G(Y)$ and \mbox{$Q = (5 + 6G(Y))^{-1}$}.

In order to avoid possible singularities in the ``troublesome term'' in the Friedman constraint \eqref{Constraint}; namely, $(2 \mathcal{P}_X - G)$, we define a new  dimensionless variable\footnote{If we had chosen to continue the computations with $y$ instead of considering the new variable $l$, we would have found divergences in the points where $l = 0$ shown in Table \ref{Table DGC}.}
\begin{equation}
l \equiv c_l \frac{\sqrt{3} H \, e^{ \lambda \phi / 2 m_\text{P}}}{m_\text{P}} \propto \frac{1}{y},
\end{equation}
with $c_l = b^{1/2}$, such that
\begin{equation}
2 \mathcal{P}_X - G = - 1 + 3 l^2 x^2.
\end{equation}
The evolution equation for this new variable is
\begin{equation} \label{l eq}
l' = \frac{\sqrt{6}}{2} \lambda l x - l (q + 1).
\end{equation}

The fixed points are now obtained by setting all the derivatives of the variables to zero in the autonomous set, and their stability can be investigated by perturbing the autonomous set around them, considering the variable $l$ instead of $y$. We mainly concentrate in the properties of the dark energy dominated points, while the other points are cataloged as viable cosmological points depending on these dark energy points. For that reason, the radiation and matter points are gathered in Table \ref{Table DGC}.

\begin{center}
\begin{table*}
\begin{centering}
\begin{tabular}{cccccccc}
\toprule 
Fixed Point & $x$ & $l$ & $z$ & $\Omega_{r}$ & $\Omega_{m}$ & $\Sigma$ & $q$ \tabularnewline
\midrule
\midrule 
(\emph{R-1}) & 0 & 0 & 0 & 1 & 0 & 0 & 1 \tabularnewline
\midrule
(\emph{M-1}) & 0 & 0 & 0 & 0 & 1 & 0 & 1/2 \tabularnewline
\midrule
(\emph{M-2}) & $\frac{\sqrt{6}}{2 \lambda}$ & $\sqrt{\frac{2}{3}} \lambda$ & 0 & 0 & $1 - \frac{3}{\lambda^2}$ & 0 & 1/2 \tabularnewline
\midrule
(\emph{M-3}) & $\frac{\sqrt{6}}{2 \lambda}$ & $\lambda \sqrt{\frac{2}{3} + \frac{1}{6} \lambda \mu - \mu^2}$ & $\sqrt{\frac{9 \mu}{8\lambda} - \frac{3}{16}}$ & 0 & $\frac{3( -8 + 3\lambda^2 - 4 \lambda \mu + 12 \mu^2)}{8 \lambda^2}$ & $\frac{6 \mu - \lambda}{4 \lambda}$ & 1/2 \tabularnewline
\midrule
(\emph{M-4}) & $\frac{\sqrt{\frac{3}{2}}\mu}{6 \mu^2 - 4}$ & $0$ & $\frac{1}{2}\sqrt{\frac{3}{6\mu^2 - 4}}$ & 0 & $\frac{9 - 12 \mu^2}{8 - \mu^2}$ & $\frac{1}{6\mu^2 - 4}$ & 1/2 \tabularnewline
\midrule
\bottomrule
\end{tabular}
\par\end{centering}
\caption{Fixed points for the DGC model. The points are labelled according to the cosmological regime as (\emph{R- }) (radiation) and (\emph{M- }) (matter).}
\label{Table DGC}
\end{table*}
\par\end{center}

\subsection{Dark Energy Fixed Points}

\begin{itemize}
\item \textbf{(\emph{DE-1}): Isotropic Dark Energy}
\end{itemize}
This is the usual isotropic dark energy fixed point in the DGC model, in which the variables take the following values:
\begin{equation*}
x = \frac{1}{2} \left( - \sqrt{\frac{3}{2}} \lambda +  \sqrt{8 + \frac{3 \lambda^2}{2}} \right), \ z = 0, \ \Sigma = 0, \ \Omega_r = 0, 
\end{equation*}
\begin{equation}
l = \sqrt{\frac{1}{4} + \frac{3 \lambda^2}{16} + \frac{3 \lambda^4}{128} + \frac{\lambda (16 + 3 \lambda^2)^{3/2}}{128 \sqrt{3}}},
\end{equation}
where we can see that the density of the vector vanishes, and then the shear is zero. The effective equation of state is 
\begin{equation}
w_{\text{eff}} = w_\text{DE} = - 1 + \frac{\lambda}{6} \left( -3 \lambda + \sqrt{48 + 9 \lambda^2} \right),
\end{equation}
and then, this point is an accelerated solution for $\lambda^2 < 2 / 3$, as expected \cite{Amendola:2015ksp}. However, as in the canonical scalar field case, the stability of the point change with the introduction of the coupling. The eigenvalues of $\mathbb{M}$ in the point are
\begin{equation*}
\eta_{1, 2} = - \frac{3}{2}(1 - w_\text{DE}), \quad \eta_3 = - 1 + 3 w_\text{DE}, \quad \eta_4 = 3 w_\text{DE},
\end{equation*}
\begin{equation}
\eta_5 = - 2 + \frac{3}{2}(1 + w_\text{DE}) (\lambda + 2 \mu).
\end{equation}
For $\lambda \geq 0$ these eigenvalues are all negative whenever
\begin{equation} \label{DGC DE1}
\lambda^2 < 2/3 \quad \land \quad \mu \leq - \frac{\lambda}{4} + \frac{\sqrt{16 + 3 \lambda^2}}{4 \sqrt{3}}.
\end{equation}
Note that $\lambda \approx 0$ in order to have $w_\text{DE} \approx -1$, implying $\mu \lesssim 1 / \sqrt{3}$. \\

\begin{itemize}
\item \textbf{(\emph{DE-2}): Anisotropic Dark Energy}
\end{itemize}

\begin{figure}[t!]
\includegraphics[width=0.92\linewidth]{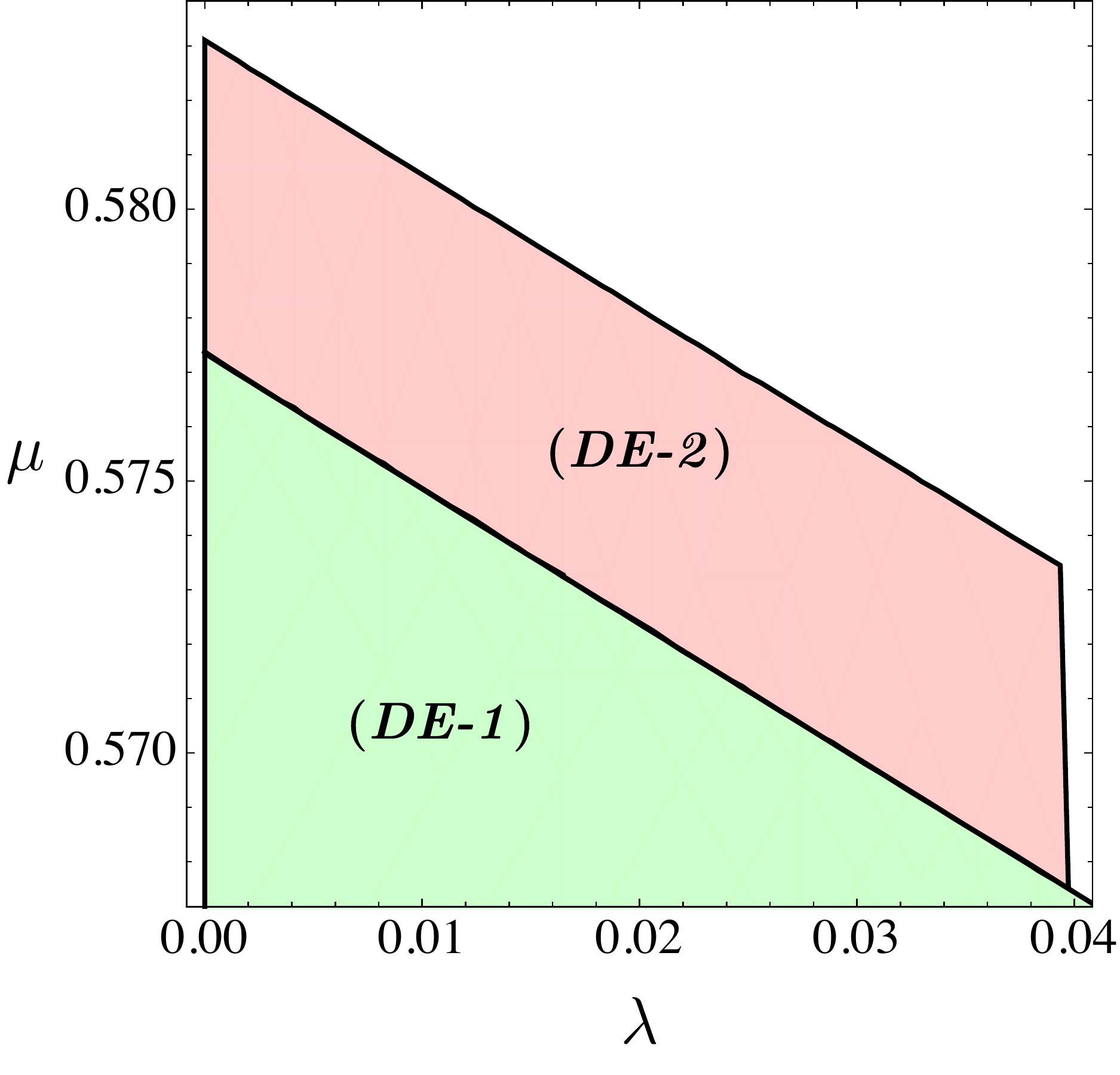}    
\caption{Stability regions for the dark energy dominated points (\emph{DE-1}) and (\emph{DE-2}) in the DGC model. Each color represents a ($\lambda, \mu$) parameter region where the indicated fixed point is an attractor with \mbox{$-1 \leq w_\text{DE} \leq -0.95$} and the shear is small \mbox{$\Sigma \leq 10^{-2}$}. The existence regions are separated by the bifurcation curve \mbox{$4 \mu = - \lambda + \sqrt{\lambda^2 + 16/3}$}.}
\label{RegionsDGC}
\end{figure}

The values of the variables in this point are
\begin{align*}
x &= 2 \sqrt{6} \frac{2\lambda + 2\mu - \sqrt{8 + \lambda^2 - 4 \lambda \mu - 8 \mu^2}}{- 8 + 3 \lambda^2 + 12 \lambda \mu + 12 \mu^2}, \\
\Sigma &= \frac{8 - 3(\lambda + 2 \mu) \left(-\lambda + \sqrt{8 + \lambda^2 - 4 \lambda \mu - 8 \mu^2}\right)}{- 8 + 3(\lambda + 2 \mu)^2}, 
\end{align*}
\begin{widetext}
\begin{align} 
l &= \frac{1}{8 \sqrt{6}} \Big\{ 9\lambda^4 - 10\lambda^3 \mu + \lambda^2 (80 - 124 \mu^2) - 8\lambda \mu ( 19\mu^2 - 14) + 32(\mu^4 + \mu^2 - 2) \nonumber \\
 &+ \sqrt{8 + \lambda^2 - 4\lambda \mu - 8 \mu^2} \left[9 \lambda^3 + 10 \lambda^2 \mu + \lambda (40 - 36 \mu^2) + 8\mu(6 - 5 \mu^2)\right)\Big\}^{1/2} \nonumber \\
 z &= \frac{1}{3(\lambda + 2 \mu^2)} \Bigg\{-96 - \frac{3}{2}\left[-9 \lambda^4 + 30 \lambda^3 \mu + 96 \mu^2 + 20\lambda^2 (4 + 9\mu^2) + 8 \lambda \mu (21 \mu^2 - 10) \right] \nonumber \\
 &+ \sqrt{8 + \lambda^2 - 4\lambda \mu - 8 \mu^2} \left[ \frac{27}{2} \lambda^3 + 72 \mu - 9 \lambda^2 \mu - 108 \mu^3 + \lambda(60 - 126 \mu^2) \right] \Bigg\}^{1/2}, \label{DGC DE2 variables}
\end{align}
and 
\begin{equation}
w_\text{DE} = -1 + 4 \lambda \frac{2\lambda^2 + 2 \mu^2 - \sqrt{8 + \lambda^2 - 4\lambda \mu - 8 \mu^2}}{- 8 + 3(\lambda + 2\mu)^2}.
\end{equation}
\end{widetext}
An equation of state of dark energy near to $-1$ requires $\lambda \sim 0$. The region of existence where this point exists as an accelerated solution, i.e. $x^2$, $l^2$, $z^2$, $\Sigma^2 > 0$, and\footnote{Note that we concentrate in no-ghost solutions. In the canonical scalar field case, it was not necessary to consider this explicitly since the  equation of state of dark energy \eqref{wDEQ} cannot cross the phantom line.} $-1 \leq w_\text{DE} < - 1/3$, is given by $\lambda < 2\sqrt{2 /39}$ and
\begin{equation} \label{DGC DE2}
-\frac{\lambda}{4} + \frac{\sqrt{16 + 3 \lambda^2}}{4 \sqrt{3}} \leq \mu \leq -\frac{\lambda}{4} + \frac{\sqrt{16 + 3 \lambda^2}}{4}.
\end{equation}
A small shear, as mandatory by observations \mbox{($| \Sigma_0 | \leq 10^{-3}$)}, requires $\mu$ to be near to its lower bound; i.e. $1 / \sqrt{3} \lesssim \mu$, which is exactly the opposite of the condition on $\mu$ for the isotropic point (\emph{DE-1}) in Eq. \eqref{DGC DE1}. As in the canonical scalar field case, this means that when the anisotropic point (\emph{DE-2}) is an accelerated solution, the isotropic point (\emph{DE-1}) is a saddle. In this case, it was not possible to obtain an analytic expression for the eigenvalues of $\mathbb{M}$ due to the complicated algebraic terms for the variables. However, we can investigate them numerically. In Fig. \ref{RegionsDGC}, we can see that the two possible dark energy points are separated by a bifurcation curve, and that the region where the anisotropic exists as an attractor is small in comparison with the region for the isotropic point. This implies that a viable cosmological evolution with a strong coupling regime between the scalar field and the vector field is not possible. Hence for instance, if we choose parameters such that (\emph{DE-2}) is the attractor, the system will spend a large amount of time around the saddle isotropic (\emph{DE-1}), reaching the anisotropic solution in the far-far future. In the numerical analysis section for this model, we corroborate that this qualitative behavior indeed occurs.

\subsection{Matter Fixed Points}

The point (\emph{M-1}) is the usual saddle isotropic point with no contributions of radiation or dark energy, while (\emph{M-2}) is the usual isotropic matter-dark energy scaling where the CMB bound on early dark energy \cite{Aghanim:2018eyx} $\Omega_\text{DE} < 0.02$ requires $\lambda > 12.25$. The point (\emph{M-3}) is a scaling solution with non-negligible shear. However, a small shear requires $\lambda \approx 6 \mu$ which is a kind of fine-tuning, and the CMB bound gives $\lambda > 11.37$. The point (\emph{M-4}) is another anisotropic scaling solution, which the CMB bound gives $\mu > 2.2$. Since the existence and stability of the anisotropic dark energy point require a small $\lambda$ and $\mu \sim 1/\sqrt{3}$, the points \mbox{(\emph{M-2})}, (\emph{M-3}), and (\emph{M-4}) are ruled out as possible matter dominated points, and thus the isotropic point (\emph{M-1}) will be dynamically selected.

\subsection{Radiation Fixed Points}

The point (\emph{R-1}) is the usual saddle isotropic point with no contributions of matter or dark energy. The system did not provide any radiation-dark energy scaling point.

From this analysis we conclude that the cosmological trajectory of this model will be given by

\begin{center}
(\emph{R-1}) \ $\rightarrow$ \ (\emph{M-1}) \ $\rightarrow$ \ (\emph{DE-1})/(\emph{DE-2}).
\end{center}

\subsection{Cosmological evolution} \label{CE}

In order to solve numerically the autonomous set in Eqs. \eqref{x eq}-\eqref{r eq} and \eqref{l eq}, we have chosen the following initial conditions:
\begin{equation*}
x_i = 4 \times 10^{-12}, \quad l_i = 3.5 \times 10^{12}, \quad z_i = 10^{-8},
\end{equation*}
\begin{equation} \label{Q init cond}
\Sigma_i = 0, \quad \Omega_{r_i} = 0.99987,
\end{equation}
as initial conditions at redshift $z_r = 2.41 \times 10^7$, and we have integrated the system up to $z_r \rightarrow -1$. Since $\Sigma_i = 0$, the starting point for any cosmological trajectory is the isotropic radiation point (\emph{R-1}). We will choose parameters $\lambda$ and $\mu$ such that the attractor point is given by (\emph{DE-2}). In particular, we have chosen $\lambda = 10^{-3}$, and $\mu = 0.58$ from the lower bound in Eq. \eqref{DGC DE2}. In this case, the numerical values  of the eigenvalues of $\mathbb{M}$ at this point are approximately
\begin{equation}
(-3.98, -2.99, -2.98, -2.98, -0.005),
\end{equation}
corroborating that (\emph{DE-2}) is an attractor. \\

In this case, we do not plot the expansion history of the Universe since it is very similar of that plotted in Fig. \ref{QAbundances}. But we do mention that the dark energy-dominance ($\Omega_\text{DE} \approx 1$ and $w_\text{eff} < -1/3$) starts from $z_r = 0.3$ and on into the future, agreeing with the results given by the dynamical system analysis, i.e. (\emph{DE-2}) is an attractor, which is further supported by the fact that the values of $x$, $l$, $z$, $\Sigma$ and $w_\text{DE}$ are those predicted by the dynamical system. Explicitly, $x \approx 1.41361$, $l \approx 0.499653$, $z \approx -0.0868072$, $\Sigma \approx 0.005$, and $w_\text{DE} \approx -0.998846$ in the far future ($z_r \rightarrow -1$), which are consistent with the values computed from the (\emph{DE-2}) in Eq. \eqref{DGC DE2 variables}. 

\begin{figure}[t!]
\includegraphics[width=0.92\linewidth]{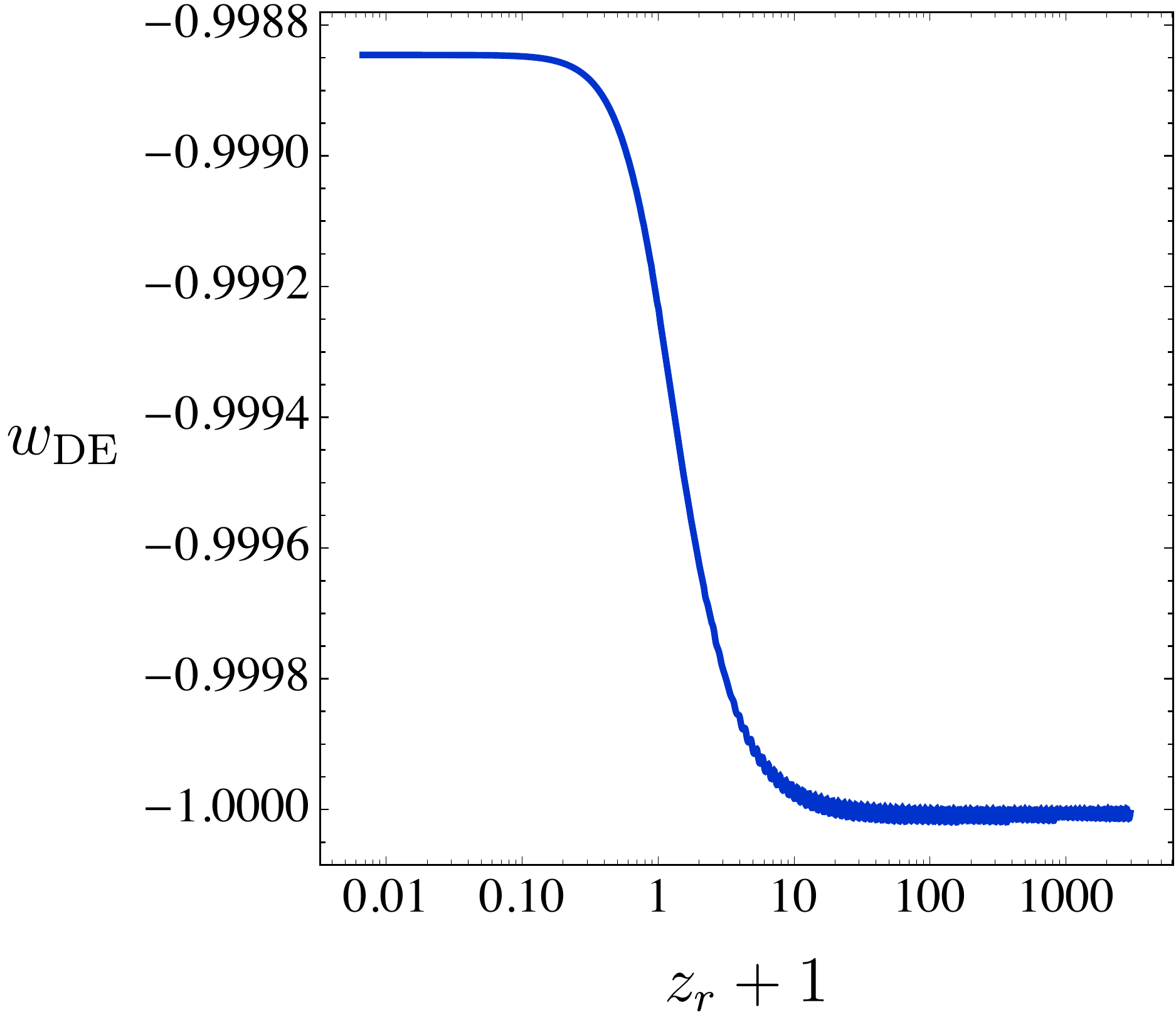}    
\caption{Evolution of the equation of state of dark energy $w_\text{DE}$ at late times. We clearly see that $w_\text{DE}$ grows from $-1$ until it gets the value predicted analytically.}
\label{DGC wDE}
\end{figure}

In Fig. \ref{DGC wDE}, we show the late-time evolution of the equation of state of dark energy but we do not plot the evolution of the shear since it is very close to zero; i.e. the shear is negligible in the expansion history of the Universe and becomes relevant only when\footnote{In this particular numerical solution, $\Sigma$ grows appreciably around $z_r = - 1 + 10^{-\infty}$. We would clarify that this extremely small value indeed corresponds to a large number of $e$-folds in the future, such that instabilities in the numerical solutions are not present. For example, since $N = - \text{ln}\, (1 + z_r)$, then $N \approx 2100$, in this case.} $z_r \rightarrow -1$. This is so because the anisotropic point is very near to the isotropic point in the parameter space ($\lambda, \mu$), and hence the system spend much time around the saddle (\emph{DE-1}), confirming the qualitative behavior expected from the dynamical analysis. Therefore, the cosmological evolution is given by
\begin{center}
(\emph{R-1}) \ $\rightarrow$ \ (\emph{M-1}) \ $\rightarrow$ \ (\emph{DE-1}) \ $\rightarrow$ \ (\emph{DE-2}),
\end{center}
but (\emph{DE-2}) is reached only in the far future.

\section{Anisotropic Dirac Born Infeld Field} \label{DBI Section}

At last, we study the Dirac Born Infeld (DBI) model whose Lagrangian reads
\begin{equation}
\mathcal{P} (\phi, X) \equiv - \frac{1}{h(\phi)} \sqrt{1 - 2 h(\phi) X} + \frac{1}{h(\phi)} - V(\phi),
\end{equation}
and thus we have $\mathcal{P}_X \equiv \gamma = ( 1 - 2 h X )^{-1}$, while the function characterizing the Lagrangian is
\begin{equation} \label{DBI G}
G(Y) = - \left( \frac{m^4}{Y} \right) \frac{1}{\gamma} - \left( \frac{M^4}{Y} \right),
\end{equation}
where $h = h (\phi)$ and $V = V(\phi)$ are functions of the scalar field $\phi$, and $m$ and $M$ are mass coefficients. The function $G(Y)$ so defined requires
\begin{align}
h (\phi) &\equiv \frac{1}{m^4} e^{\lambda \phi / m_\text{P}}, \\
V(\phi) &\equiv m^4(1 + \eta) e^{- \lambda \phi / m_\text{P}},
\end{align}
where we have defined 
\begin{equation}
m^4 \equiv c_m^2 m_\text{P}^4, \ M^4 \equiv c_M^2 m_\text{P}^4, \ \eta \equiv \frac{c_M^2}{c_m^2},
\end{equation}
with $c_m$ and $c_M$ constants.

Possible singularities in the ``troublesome term'' $x^2(2 \mathcal{P}_X - G)$ can be avoided if we use both variables $y$ and $l$,
such that
\begin{equation}
x^2 ( 2 \mathcal{P}_X - G ) = \frac{2 \eta \, x^2 - y^2 \left( \eta + \sqrt{1 - 2 l^2 x^2} \right)}{- 1 + 2 l^2 x^2}.
\end{equation}
The fixed points should be obtained by setting all the derivatives of the variables equal to zero in the autonomous set composed by all the variables, i.e. Eqs. \eqref{x eq} - \eqref{r eq} and Eqs. \eqref{y eq} and \eqref{l eq}. However, this task is analytically impossible for general values of the parameters $\lambda$, $\mu$, and $\eta$ \cite{Ohashi:2013pca}. Instead of that, we look for the radiation, matter, and dark energy points for apart. We begin by finding the dark energy dominated points.

\begin{center}
\begin{table*}
\begin{centering}
\begin{tabular}{ccccccccc}
\toprule 
Fixed Point & $x$ & $y$ & $l$ & $z$ & $\Omega_{r}$ & $\Omega_{m}$ & $\Sigma$ & $w_{\text{eff}}$ \tabularnewline
\midrule
\midrule 
\emph{R-1} & 0 & 0 & 0 & 0 & 1 & 0 & 0 & 1/3 \tabularnewline
\midrule
\emph{R-2} & $\frac{2 \sqrt{6}}{3 \lambda}$ & $\frac{2}{\sqrt{3} \lambda} \sqrt{\frac{4\eta^2 - \eta - 3}{\eta^2 - 1}}$ & 0 & 0 & $1 - \frac{4}{\lambda^2}$ & 0 & 0 & 1/3 \tabularnewline
\midrule
\emph{R-3} & $\frac{2 \sqrt{6}}{3 \lambda}$ & $\frac{\sqrt{6}}{3 \lambda} \sqrt{\frac{6 + 8\eta + 3\mu^2}{\eta + 1}}$ & 0 & $\sqrt{\frac{\mu}{\lambda}}$ & $1 - \frac{4 + \lambda \mu + 6 \mu^2}{\lambda^2}$ & 0 & $\frac{2\mu}{\lambda}$ & 1/3 \tabularnewline
\midrule
\emph{M-1} & 0 & 0 & 0 & 0 & 0 & 1 & 0 & 0 \tabularnewline
\tabularnewline
\midrule
\bottomrule
\end{tabular}
\par\end{centering}
\caption{Fixed points for the DBI model. The points are labelled according to the cosmological regime as \emph{R-} (radiation) and \emph{M}- (matter).}
\label{Table DBI}
\end{table*}
\par\end{center}

\subsection{Dark Energy Fixed Points}

We follow the procedure of Ref. \cite{Ohashi:2013pca} in order to find the dark energy dominated points. At very late times, dark energy dominates and the contributions of radiation and matter to the total density are negligible. Therefore, we consider $\Omega_r = \Omega_m = 0$, and then, from the constraint in Eq. \eqref{Constraint}, we can write the variable $z$ in terms of $x$, $y$, and $\Sigma$. The fixed points of this simplified system are found by setting $x' = 0$, $\Sigma' = 0$, and $y' = 0$ in Eqs. \eqref{x eq}, \eqref{Sigma eq}, and \eqref{y eq}. We obtain:
\begin{itemize}
\item (\emph{DE-1}) Isotropic point
\end{itemize}
\begin{equation} \label{DBI Iso vars}
\mathcal{P}_X = \frac{\lambda}{\sqrt{6} x}, \ G_1 = \frac{1}{x^2} - \frac{\lambda}{\sqrt{6} x}, \ z = 0, \ \Sigma = 0,
\end{equation}
\begin{itemize}
\item (\emph{DE-2}) Anisotropic point
\end{itemize}
\begin{align} \label{DBI Ani vars}
\mathcal{P}_X &= \frac{(\lambda + 2 \mu)\left( 2\sqrt{6} - x(\lambda + 6 \mu)\right)}{8x}, \nonumber \\
G_1 &= \frac{12 + x^2 \lambda (\lambda + 2 \mu) -  \sqrt{6} x ( 3 \lambda + 2 \mu)}{8x^2}, \nonumber \\
\Sigma &= - 1 + \frac{\sqrt{6}}{4} x (\lambda + 2 \mu).
\end{align}
In both cases we have
\begin{equation}
w_\text{DE} = -1 + \sqrt{\frac{2}{3}} \lambda x,
\end{equation}
and thus accelerated solutions obey
\begin{equation}
0 \leq x < \frac{\sqrt{6}}{3 \lambda}.
\end{equation}
In the next subsections, we will study each point apart.

\begin{itemize}
\item \textbf{(\emph{DE-1}): Isotropic Dark Energy}
\end{itemize}

In this case, it is possible to find explicit expressions for the variables $x$ and $y$. By noting that $\gamma = \mathcal{P}_X = (1 - 2 Y/m^4)^{-1/2}$ and that $Y/m^4 = x^2 / y^2$, we can use Eq. \eqref{DBI Iso vars} to write $y$ in terms of $x$ as
\begin{equation}
y = \frac{\sqrt{2} x \lambda}{\sqrt{\lambda^2 - 6 x^2}}.
\end{equation}
Now, since $G_1(Y) = \gamma - G(Y)$, where $G(Y)$ is given in Eq. \eqref{DBI G}, we get the following algebraic equation for $x$:
\begin{equation}
x^2(3 + \eta \lambda^2) + \frac{\lambda^3}{\sqrt{6}} x - \frac{1}{2} \lambda^2 = 0,
\end{equation}
from which we obtain
\begin{equation}
x = \frac{\lambda \left( - \lambda^2 \pm \sqrt{36 + 12\eta \lambda^2 + \lambda^4}\right)}{2\sqrt{6} (3 + \eta \lambda^2)}.
\end{equation}
The minus solution does not correspond to an accelerated solution, hence we only consider the solution with the plus sign. This point is an accelerated when
\begin{equation}
\lambda > 0 \quad \land \quad \eta > \frac{\lambda^4 - 12}{4 \lambda^2}.
\end{equation}
Note that, with respect to the other models studied, in this case we can get $w_\text{DE} \approx - 1 $ for $\lambda = \mathcal{O}(10)$ and large $\eta$. For instance, if $\lambda = 1$, $\eta > 3297$ for $- 1 \leq w_\text{DE} < -0.99$.

\begin{itemize}
\item \textbf{(\emph{DE-2}): Anisotropic Dark Energy}
\end{itemize}

In the isotropic case, the variable $x$ can be obtained from a third order algebraic equation which has analytic solutions. However, in the anisotropic case, we get a fourth order algebraic equation for $x$ which is \emph{not} analytically solvable for general values of $\lambda$, $\eta$, and $\mu$. Therefore, it is impossible to find explicit expressions for the variables. However, we do find bounds for them in order to discriminate cosmologically viable solutions.

Using Eq. \eqref{DBI Ani vars} and the Friedman constrain, we get $z^2$ and
\begin{equation}
z = \frac{\sqrt{(\sqrt{6} - x \lambda) \left(- 2 \sqrt{6} + 3x (\lambda + 2 \mu) \right)}}{2 \sqrt{2}},
\end{equation}
which takes real values whenever
\begin{equation}
\frac{2 \sqrt{6}}{3(\lambda + 2 \mu)} < x < \frac{\sqrt{6}}{3 \lambda}.
\end{equation}
A tighter bound can be obtained by noting that in order to $\mathcal{P}_X$ be positive, $x$ must be
\begin{equation}
0 < x < \frac{2 \sqrt{6}}{\lambda + 6 \mu},
\end{equation} 
and thus
\begin{equation} \label{x bound}
\frac{2 \sqrt{6}}{3(\lambda + 2 \mu)} < x < \frac{2 \sqrt{6}}{\lambda + 6 \mu}.
\end{equation}
A cosmologically viable anisotropic accelerated solution must fulfill that $w_\text{DE} \approx - 1$ and $| \Sigma_0 | \leq 10^{-3}$ \cite{Aghanim:2018eyx, Campanelli:2010zx, Amirhashchi:2018nxl}. We can see that, if we take $x$ equal to its lower bound then
\begin{equation}
w_\text{DE} = - 1 + \frac{4 \lambda}{3 (\lambda + 2 \mu)}, \quad \Sigma = 0.
\end{equation}
Therefore, we have to choose the parameters $\lambda$ and $\mu$ such that $x$ approximate its lower bound, and also $\mu \gg \lambda$. Now, using $G_1$ from Eq. \eqref{DBI Ani vars} and having in mind that $G_1 = \gamma - G$ we get the compatibility conditions
\begin{equation} \label{DBI mu}
\mu > \frac{2 \sqrt{\lambda^4 + 12 \eta \lambda^2 + 36} - \lambda^2}{6\lambda},
\end{equation}
\begin{equation} \label{DBI lambda}
\lambda < \sqrt{2\eta + 2\sqrt{3 + \eta^2}},
\end{equation}
where we assumed that $x$ is equal to the lower bound of Eq. \eqref{x bound}. These results assure that $\mu \gg \lambda$. Note that this procedure considers a general $k$-essence field. Therefore, we can say that a $k$-essence field coupled to a vector field can give rise to anisotropic accelerated solutions, although the full available  parameter space does depend on the particular Lagrangian. Nonetheless, we want to stress that we do not used this approach from the beginning since it is only useful for the dark energy dominated points corresponding to non-zero values of $x$ and $y$, and hence the radiation or matter fixed points are not considered. 

\subsection{Radiation and Matter Fixed Points}

Firstly, we assume that radiation is negligible and that the Universe is dust dominated; i.e. $\Omega_r = 0$ and $q = 1/2$ ($w_\text{eff} \approx 0$). Then we get the matter dominated point shown in Table \ref{Table DBI}. This is the usual isotropic matter point with negligible contributions of dark energy and radiation. This procedure did not returned any scaling matter-dark energy point. Then, we assumed that the Universe is radiation dominated: $q = 1$ ($w_\text{eff} \approx 1/3$). The resulting radiation dominated points are shown in Table \ref{Table DBI}. The point \mbox{(\emph{R-1})} is the usual isotropic radiation dominated point. The point (\emph{R-2}) is a scaling solution which requires $\lambda > 9.43$ in order to obey the BBN bound on early dark energy. The anisotropic scaling point \mbox{(\emph{R-3})} also needs $\lambda > 9.43$ in order to follow this bound, but the coupling $\mu$ has to be very small for a small shear during the radiation epoch. 
Therefore, also in this model, the cosmological trajectory would be 
\begin{center}
\mbox{(\emph{R-1}) \ $\rightarrow$ \ (\emph{M-1}) \ $\rightarrow$ \ (\emph{DE-1}) \ $\rightarrow$ \ (\emph{DE-2})}.
\end{center}

\subsection{Cosmological evolution} \label{CE}
In order to solve numerically the autonomous set in Eqs. \eqref{x eq}-\eqref{r eq}, and \eqref{y eq}, we have chosen the following initial conditions:
\begin{equation*} 
x_i = 10^{-17}, \quad y_i = 1.6 \times 10^{-16}, \quad z_i = 10^{-12},
\end{equation*}
\begin{equation} \label{DBI init cond}
\Sigma_i = 0, \quad \Omega_{r_i} = 0.99995,
\end{equation}
as initial conditions at redshift $z_r = 6.56 \times 10^7$, and we have integrated the system up to $z_r \rightarrow -1$. Since $\Sigma_i = 0$, the starting point for any cosmological trajectory is the isotropic radiation point (\emph{R-1}). In particular, we have chosen for the parameters $\lambda = 1$, $\mu = 5000$, and $\eta = 15000$ such that the attractor point is given by (\emph{DE-2}), i.e. we consider a small $\lambda$ and large $\eta$ and $\mu$ assuring the existence of a strong coupling regime.  Although there are no analytical expressions for the eigenvalues of $\mathbb{M}$ neither at (\emph{DE-1}) nor at (\emph{DE-2}) for general values of the parameters, we can investigate numerically the stability of the point for specific parameters. In this case, the numerical values  of the eigenvalues of $\mathbb{M}$ at (\emph{DE-1}) are approximately
\begin{equation}
(0, -2.99, 7.95),
\end{equation}
while for (\emph{DE-2}) are
\begin{equation}
(0, -3.99, -2.99),
\end{equation}
corroborating that (\emph{DE-2}) is ``more stable'' than (\emph{DE-1}). \\

In this case, we also do not plot the expansion history of the Universe since it is similar of that plotted in Fig. \ref{QAbundances}. But we do mention that the system evolves as predicted in the dynamical system analysis, i.e. (\emph{DE-2}) is an attractor. This is further supported by noting that if \mbox{(\emph{DE-2})} is the attractor, the variables take approximately the following values: \mbox{$x \approx 1.63283 \times 10^{-3}$} and $w_\text{DE} \approx -0.999867$. The numerical solution shows that in the far future \mbox{($z_r = -1 + e^{-50}$)}, \mbox{$x \approx 1.63294 \times 10^{-3}$}, \mbox{$w_\text{DE} \approx -0.999867$}, which are consistent with the values predicted for (\emph{DE-2}), moreover $\Sigma \approx 6.66 \times 10^{-5}$.\footnote{In the case that (\emph{DE-1}) is the attractor, the values of the variables would be $x \approx - 5.8 \times 10^{-3}$, $w_\text{DE} \approx -1.005$, and $\Sigma = 0$ of course.}

\begin{figure}[t!]
\includegraphics[width=0.92\linewidth]{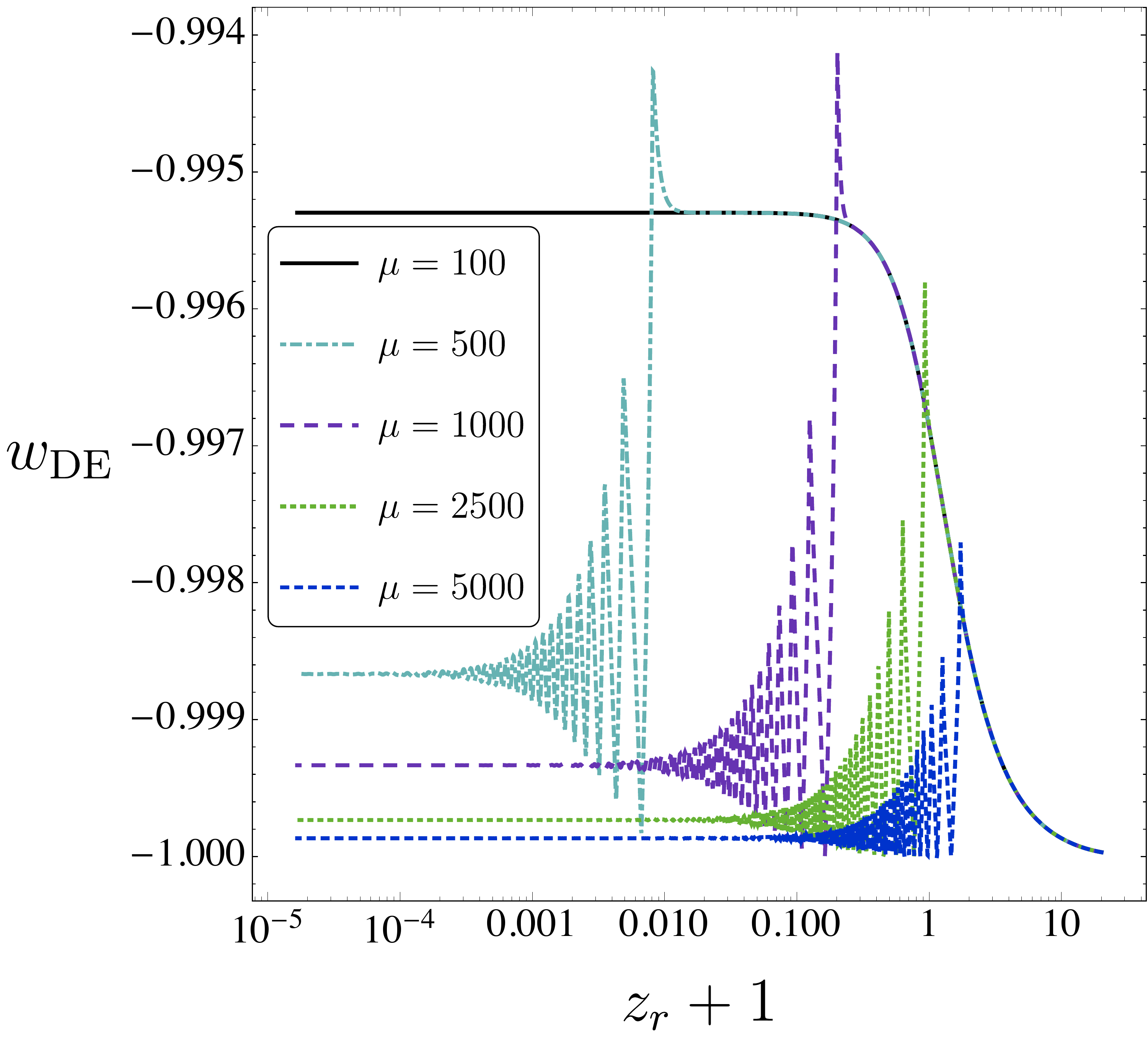}    
\caption{Evolution of the equation of state of dark energy $w_\text{DE}$ around $z_r = 0$ for different values of the parameter $\mu$, while $\lambda$ and $\eta$ are fixed and the initial conditions are the same given in Eq. \eqref{DBI init cond}. }
\label{DBIOscwDE}
\end{figure}

\begin{figure}[t!]
\includegraphics[width=0.92\linewidth]{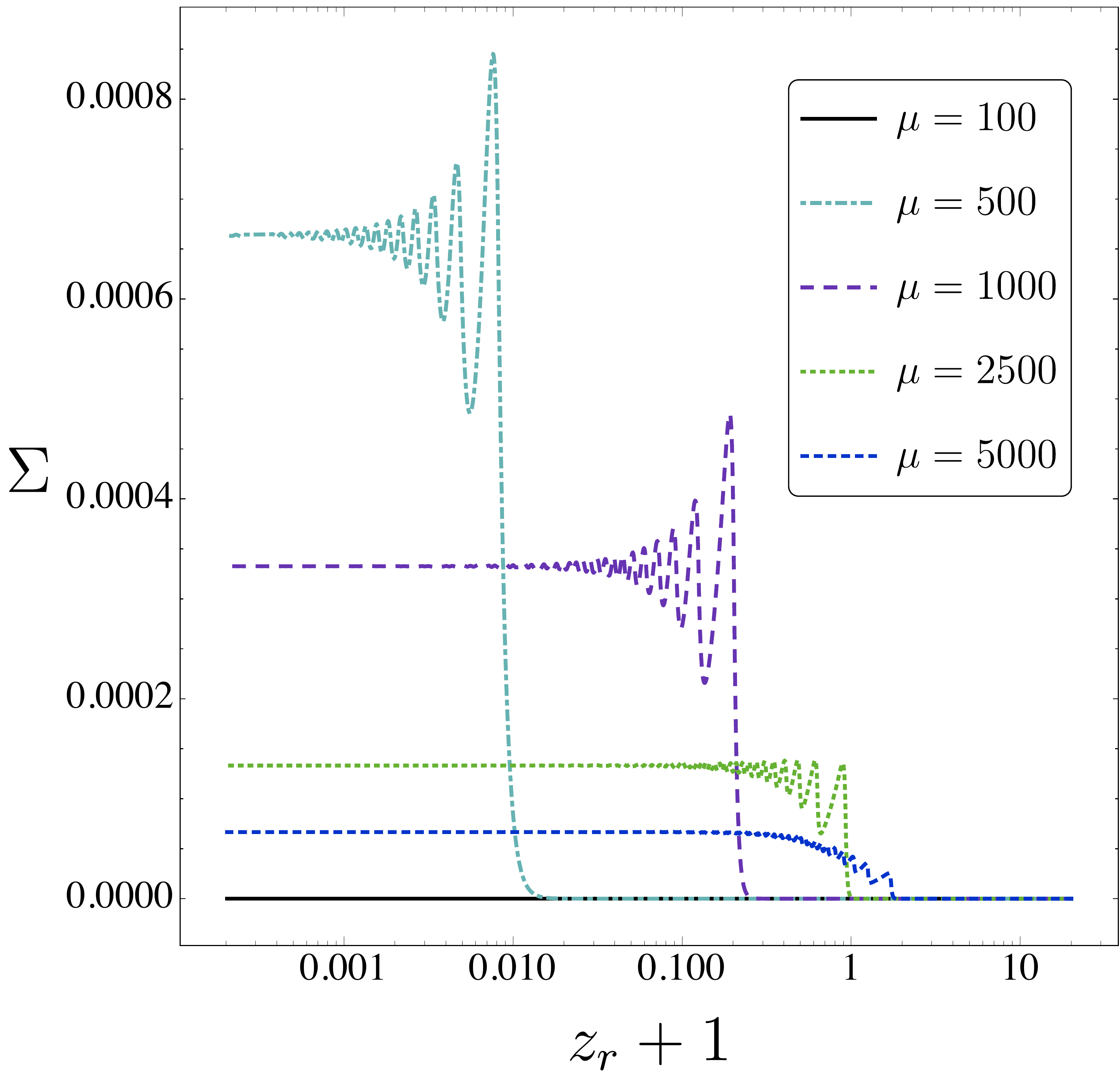}    
\caption{Evolution of the shear $\Sigma$ around $z_r = 0$ for different values of the parameter $\mu$, while $\lambda$ and $\eta$ are fixed and the initial conditions are the same given in Eq. \eqref{DBI init cond}.}
\label{DBIOscShear}
\end{figure}

In Figs. \ref{DBIOscwDE} and \ref{DBIOscShear}, we show the late-time evolution of the equation of state of dark energy and the shear. As in the canonical scalar field case, they perform quickly oscillations at late times when the parameter $\mu$ is large enough to the vector field contribute significantly to the total density of the Universe (see Appendix \ref{App}). In conclusion, the cosmological trajectory of the model is also given by
\begin{center}
(\emph{R-1}) \ $\rightarrow$ \ (\emph{M-1}) \ $\rightarrow$ \ (\emph{DE-1}) \ $\rightarrow$ \ (\emph{DE-2}).
\end{center}

\section{Conclusions} \label{Conclusions}

In this work, we have studied the background cosmological evolution of a $k$-essence field coupled to the canonical kinetic term of an abelian vector field, extending to late times the work of Ref. \cite{Ohashi:2013pca} where inflationary solutions were found. We assumed three specific realizations of the $k$-essence field, namely: the dilatonic ghost condensate, the Dirac Born Infeld field, and the canonical scalar field for completeness.

By using a dynamical system approach, we were able to show that each of these models has an anisotropic dark energy dominated point which can be an attractor. The dynamical system also allowed us to find the available space parameter for each anisotropic solution to be cosmologically viable. For the quintessence field, an equation of state of dark energy near to $-1$ and a small shear as required by observations \cite{Aghanim:2018eyx, Campanelli:2010zx, Amirhashchi:2018nxl}, need $\lambda \ll 0$ and $\mu \gg \lambda$ and $\mu \gg 1$ (see the bounds in Eqs. \eqref{QDE2} and Fig. \ref{Regions}). For the DGC model, the requirements are $\lambda \ll 1$ and $\mu \gtrsim 1/\sqrt{3}$ (see the bounds in Eq. \eqref{DGC DE2} and Fig. \ref{RegionsDGC}), while for the DBI field we found that $\lambda$ can be greater than 1 ($\lambda = \mathcal{O}(10)$, for instance) given that $\eta, \mu \gg 1$ [see the bounds in Eqs. \eqref{DBI mu} and \eqref{DBI lambda}].

We solved numerically the autonomous set of equations corresponding to each model, verifying that a correct expansion history is reproduced, and that the anisotropic dark energy dominated points (\emph{DE-2}) are indeed attractors of the systems. For every case studied, we set the initial conditions in the deep radiation epoch [$z_r = \mathcal{O}(10^{7})$] and assumed a zero initial shear ($\Sigma_i = 0$). For the quintessence and the DBI field, we found that $w_\text{DE}$ and $\Sigma$ oscillate at late times until they reach their values in the attractor (\emph{DE-2}). In Appendix \ref{App}, we show that for parameters $\lambda \ll 0$ and $\mu \gg \lambda$ and $\mu \gg 1$, the system is underdamped, making possible to see these oscillations around nowadays. In the uncoupled case ($\mu = 0$), these oscillations are overdamped and thus it is extremely hard to see them. We also estimated the value of $\mu$ in order to get critically damped oscillations. However, these critically damped oscillations could not be see around $z_r = 0$, since they occur in the far far future [in other words, the value of $\mu$ is not enough to the cosmological trajectory escape quickly from the isotropic point (\emph{DE-1})]. On the other hand, for the DGC model we found that the shear today (and for the near future) would be effectively zero, because the isotropic and anisotropic point are very near in the phase space. This implies that the cosmological trajectories will spend at large amount of time around the isotropic point before reach the anisotropic attractor. This also implies that no oscillations of the fields are expected to be seen. Anyway, the three concrete models have anisotropic accelerated attractors and the general cosmological trajectory can be
\begin{center}
\mbox{(\emph{R-1}) \ $\rightarrow$ \ (\emph{M-1}) \ $\rightarrow$ \ (\emph{DE-1}) \ $\rightarrow$ \ (\emph{DE-2})},
\end{center}
changing in the time at which the corresponding attractor is reached.

The non-negligible shear and the oscillatory behavior of the fields, $w_\text{DE}$, and $\Sigma$, are notorious properties of this coupled $k$-essence model. However, the observable imprints of these models which could be compared with CMB or SN Ia measurements are yet to be done. We leave a detailed study of this subject for a future work.

\section*{Acknowledgements}
This work was supported by Patrimonio Autónomo - Fondo Nacional de Financiamiento  para  la  Ciencia,  la  Tecnología  y  la  Innovación  Francisco  José  de  Caldas  (MINCIENCIAS - COLOMBIA)  Grant  No.   110685269447  RC-80740-465-2020,  projects  69723 and 69553. 

\appendix
\section{Oscillatory Behavior of $w_\text{DE}$ and $\Sigma$} \label{App}

In this section, we will explain why the oscillatory behavior of $w_\text{DE}$ and $\Sigma$ at late times for the canonical quintessence field is presented. Similar arguments apply for the DBI model. For the quintessence field we have:
\begin{align}
\mathcal{P}_X &= 1, \quad G = 1 - c \frac{m_\text{P}^4}{X} e^{- \lambda \phi / m_\text{P}}, \nonumber \\
G_1 &= 1, \quad Q = 1, \quad \frac{f_\phi}{f} = - \frac{\mu}{m_\text{P}}.
\end{align}
Defining $V_0 \equiv c m_\text{P}^4$, and replacing the above expressions in the equation of motion for the scalar field \eqref{EoM phi}, we get
\begin{equation} \label{Eq App}
\ddot{\phi} + 3 H \dot{\phi} - \lambda \frac{V_0}{m_\text{P}} e^{-\lambda \phi / m_\text{P}} + 2 \mu \frac{\rho_A}{m_\text{P}} = 0.
\end{equation}
At late times, we can neglect the contribution of radiation, matter, and the shear to the first Friedman equation \eqref{H2 eq} and write
\begin{equation}
3 m_\text{P}^2 H^2 \approx \frac{1}{2} \dot{\phi}^2 + V_0 e^{- \lambda \phi / m_\text{P}} + \rho_A.
\end{equation}
At some time the contribution of the vector density is comparable to the kinetic energy of the scalar field, $\dot{\phi}^2 \approx 2 \rho_A$, and assuming that the exponent in the exponential is small enough, we can approximate the first Friedman equation by
\begin{equation}
3 m_\text{P}^2 H^2 \approx V_0 - V_0 \lambda \frac{\phi}{m_\text{P}} + 2 \rho_A.
\end{equation}
From the dynamical system we obtained that $y \approx 1$ at late times, then $3 m_\text{P}^2 H^2 \approx V_0$, therefore
\begin{equation}
3 \lambda H^2 \phi \approx 2 \frac{\rho_A}{m_\text{P}}.
\end{equation}
Replacing this expression in Eq. \eqref{Eq App}, and taking into account that the exponent in the exponential is small, and that at late times $w_\text{DE} \approx - 1$ implying $H \approx \text{const}$, we get
\begin{equation}
\ddot{\phi}(t) + \gamma \dot{\phi}(t) + \omega_0^2 \phi(t) = F_0,
\end{equation}
which corresponds to the equation of motion of a damped harmonic oscillator with damping $\gamma$ (which is the usual Hubble friction), frequency $\omega_0^2$,  driven by a constant force $F_0$. These terms are given by
\begin{equation}
\gamma \equiv 3 H, \quad \omega_0^2 \equiv 3 \lambda(\lambda + \mu) H^2, \quad F_0 \equiv \lambda \frac{V_0}{m_\text{P}}.
\end{equation}
Hence we see that, when the coupling parameter $\mu$ is small, the frequency of the oscillator is $\omega_0^2 \approx 3 \lambda^2 H^2$ and of course $\gamma \gg \omega_0$, given that $\lambda \approx 0$, and thus the oscillator is overdamped. That is the reason why we need a large $\mu$ to see oscillations. Even more, we can give an estimate of how large $\mu$ has to be in order to see critically damped oscillations; i.e. when $\gamma^2 = 4 \omega_0^2$. Using $\lambda = 0.1$ we get $\mu = 7.4$. We confirmed that, using the same initial conditions in Eq. \eqref{Q init cond}, a very small oscillation occurs in $w_\text{DE}$ for $\mu = 8$ at redshift $z_r \approx -1 + e^{-8}$.

\bibliography{Bibli.bib}

\end{document}